\documentclass[%
reprint,
superscriptaddress,
 amsmath,amssymb,
 aps,
]{revtex4-1}

\usepackage{graphicx}
\usepackage{dcolumn}
\usepackage{bm}
\usepackage[colorlinks=true,allcolors=blue]{hyperref}

\usepackage[utf8]{inputenc}
\usepackage[T1]{fontenc}

\usepackage{pgf,tikz}
\usepackage{textcomp} 
\usetikzlibrary{matrix, arrows, automata}
\usepackage{blkarray} 
\usepackage{esint} 
\usepackage{blindtext}
\usepackage{color}
\usepackage[normalem]{ulem} 

\usepackage{multirow,booktabs} 

\usepackage{chngcntr} 

\newcommand{\units}[1]{\,\mathrm{#1}}
\renewcommand{\t}[1]{\mathrm{\text{#1}}}

\newcommand{\Dfrac}[3][{}]{\frac{\mathrm{d}^{#1} #2}{\mathrm{d} {#3}^{#1}} }
\newcommand{\0}{{(0)}}
\newcommand{\1}{{(1)}}
\newcommand{\eps}{\varepsilon}
\renewcommand{\vec}[1]{\mathbf{#1}}
\renewcommand{\d}[1]{\,\mathrm{d}#1}

\newcommand{\chg}[2][black]{{\color{#1}#2}}
\newcommand{\chgrev}[2][red]{{\color{#1}#2}}
\newcommand{\chgrevtwo}[2][red]{{\color{#1}#2}}
\newcommand\rem[1]{}   

\renewcommand{\chgrev}[2][black]{#2}
\renewcommand{\chgrevtwo}[2][black]{{\color{#1}#2}}

\begin{document}


\title{\Large{Quantifying the impact of tissue metabolism on solute transport in feto-placental microvascular networks}}

\author{Alexander Erlich}
\affiliation{School of Mathematics, University of Manchester, Oxford Road, Manchester M13 9PL, UK}
\author{Gareth A. Nye}
\affiliation{Maternal and Fetal Health Research Centre, Division of Developmental Biology and Medicine, School of Medical Sciences, University of Manchester, Manchester Academic Health Science Centre, Manchester M13 9PL, UK}
\affiliation{\chgrevtwo{Chester Medical School, University of Chester, Chester CH1 4AR, UK}}
\author{Paul Brownbill}
\affiliation{Maternal and Fetal Health Research Centre, Division of Developmental Biology and Medicine, School of Medical Sciences, University of Manchester, Manchester Academic Health Science Centre, Manchester M13 9PL, UK}
\author{Oliver E. Jensen}
\affiliation{School of Mathematics, University of Manchester, Oxford Road, Manchester M13 9PL, UK}
\author{Igor L. Chernyavsky}
\email[To whom correspondence should be addressed.\\ E-mail: ]{igor.chernyavsky@manchester.ac.uk}
\affiliation{School of Mathematics, University of Manchester, Oxford Road, Manchester M13 9PL, UK}
\affiliation{Maternal and Fetal Health Research Centre, Division of Developmental Biology and Medicine, School of Medical Sciences, University of Manchester, Manchester Academic Health Science Centre, Manchester M13 9PL, UK}


\begin{abstract}
The primary exchange units in the human placenta are terminal villi, in which fetal capillary networks are surrounded by a thin layer of villous tissue, separating fetal from maternal blood. To understand how the complex spatial structure of villi influences their function, we use an image-based theoretical model to study the effect of tissue metabolism on the transport of solutes from maternal blood into the fetal circulation. For solute that is taken up under first-order kinetics, we show that the transition between flow-limited and diffusion-limited transport depends on \chgrev{two} new dimensionless parameters defined in terms of key geometric quantities, with strong solute uptake promoting flow-limited transport conditions.  We present a simple algebraic approximation for solute uptake rate as a function of flow conditions, metabolic rate and villous geometry.  For oxygen, accounting for nonlinear kinetics using physiological parameter values, our model predicts that \chgrev{villous metabolism does not significantly impact oxygen transfer to fetal blood, although} 
the partitioning of fluxes between the villous tissue and the capillary network depends strongly on the flow regime.
\end{abstract}


\maketitle

\section{Introduction}
The human placenta is an unusual and often overlooked organ. During pregnancy, it supplies the developing fetus with all its essential nutrients, removes its waste products and has a range of additional endocrine functions \cite{Burton15}. Placental insufficiency compromises fetal growth and can have life-long impact on the later health of the individual \cite{HorikoshiMcCarthy_etal16}. As an exchange organ, the placenta's geometric structure plays a crucial role in determining its function. With three-dimensional imaging revealing placental morphological complexity in ever greater detail \cite{roth2017dynamic,perazzolo2017modelling,pearce2016image}, it is important to look at placental structure through the prism of the physical transport processes taking place within it. This allows us to identify the geometrical features that dictate transport capacity, and to characterise in quantitative terms the pathological consequences of structural abnormality.

The present study contributes to a growing literature in which mathematical and computational models have been used to provide insight into placental physiology. These studies have addressed the fetal circulation (involving networks of blood vessels confined within villous trees), the maternal circulation (involving blood flowing outside the branches of the trees, effectively through a porous medium) and solute exchange across the trophoblast barrier between them. Recent reviews are provided by \chgrev{Serov \hbox{et al.} \cite{serov2015role}},  Jensen \& Chernyavsky \cite{jensenchernyavsky2018} and Plitman Mayo \cite{mayo2018advances}. Here we focus on the primary structural exchange unit associated with the fetal circulation, namely the terminal villus: this is effectively a protruding `leaf' on a villous tree that contains an irregular network of fetal capillaries. The thin-walled villus is bathed in maternal blood, allowing dissolved gases and nutrients to pass between fetal and maternal blood. If blood flow in the capillaries is 
\chgrev{insufficient to carry available solute}
we term the transport `flow-limited'; if diffusion through villous tissue is the dominant barrier to exchange, we term the transport `diffusion-limited'.

Pearce \hbox{et al.} \cite{pearce2016image} constructed a regression equation describing maternal-to-fetal solute transport in a terminal villus, by taking a harmonic average of limiting approximations for solute fluxes valid under flow-limited and diffusion-limited conditions. As explained by \cite{jensenchernyavsky2018}, this expression is naturally expressed in terms of suitable dimensionless parameters, namely a Damk{\"{o}}hler number Da that measures solute transit time across the villous tissue due to diffusion relative to transit time through the villus due to flow, and a parameter $\mu$ that measures the relative diffusive capacities of the villous tissue and the intravillous capillary network. Erlich \hbox{et al.} \cite{erlich2018physical} added a further refinement to the regression equation and then validated it using computational simulations of four villus specimens, each having complex internal structure. The significant physical parameters in their analysis were the solute diffusivities in tissue and plasma ($D_t$ and $D_p$ respectively), the effective viscosity of blood $\eta$ (based on an assumption of Newtonian flow), a dimensionless parameter $B$ that captures the advective boost that oxygen acquires from binding to haemoglobin \cite{kasinger1981vivo}, and the imposed pressure drop $\Delta P$ driving blood through the vessel network. This analysis also revealed some of the key geometric parameters determining the transport capacity of a villus for most solutes: the flow resistance of the capillary network per unit viscosity ($\mathcal{R}/\eta$, which has dimensions of inverse volume); the total length of capillary vessels within the villus $L_c$; and a lengthscale $\mathcal{L}$ capturing the diffusive capacity of villous tissue (a normalised diffusive flux integrated over an exchange area). A key finding from \cite{erlich2018physical} is that, for the majority of physiologically relevant solutes studied, the diffusive capacity ratio $\mu\chgrev{\,=\!D_t \mathcal{L}/D_p L_c}$ was sufficiently small among all specimens studied for the effects of concentration boundary layers within capillaries to be \chgrev{a secondary factor}. Then, assuming the solute is not absorbed by villous tissue, transport was predicted to be flow-limited when $\mathrm{Da}\gg 1$ and diffusion-limited when $\mathrm{Da}\ll 1$, where
\begin{equation}
\mathrm{Da} =\frac{D_t \mathcal{L} \mathcal{R}}{B\Delta P}.
\label{eq:da}
\end{equation}
The solute flux $N$ is well approximated \cite{pearce2016image, erlich2018physical} by 
\begin{equation}
N=\frac{N_{\mathrm{max}}}{1\,+\,\mathrm{Da} \chgrev{\,+\,\mathrm{Da_F}^{1/3}}},
\label{eq:regression-classical}
\end{equation} 
where $N_{\max}=D_\text{t} \Delta c \mathcal{L}$ represents the maximum diffusive capacity of the villus under a solute concentration difference $\Delta c$ between maternal and fetal blood.   \chgrev{Setting aside the term involving $\mathrm{Da_F}\equiv \mu^2\mathrm{Da}/166.4$ (a correction accounting for concentration boundary layers), (\ref{eq:regression-classical}) captures the transition from flow-limited transport ($N\approx N_{\max}/\mathrm{Da}=B(\Delta P/\mathcal{R})\Delta c $) to diffusion-limited transport ($N\approx N_{\max}$) as flow strength (Da$^{-1}$) increases from low to high values.}  The simple expressions in (\ref{eq:da},\ref{eq:regression-classical}) show how physical processes and villous geometry together influence solute transfer. In particular, they demonstrate how, for given flow conditions, different solutes can have widely varying values of Da (through differing values of $D_t/B$), implying that flow-limited and diffusion-limited transport may take place simultaneously \chgrevtwo{in the same villus}.

This approach can be used to understand the transport of solutes that pass passively through villous tissue. For some solutes, however, the \chg{situation} is not so simple, either because active transport is required (in the case of amino acids \cite{Sibley_etal18}) or because the solute is absorbed by villous tissue. We consider such solutes here, focusing \textit{inter alia} on oxygen, a proportion of which can be taken up by villous tissue before reaching fetal blood \cite{Carter00,Schneider00}. Our primary goal is to refine the estimate of solute transfer $N$ to account for this uptake. We use simulations to compute the transfer rate in terminal villi recovered from imaging using confocal microscopy.  To describe uptake \chgrev{of a generic solute} under linear (first-order) kinetics, we introduce a kinetic parameter $\alpha$ that describes the uptake rate by villous tissue, and then present \chgrev{a modified} version of (\ref{eq:regression-classical}) that expresses uptake in terms of Da and $\alpha$. We identify \chgrev{two} new dimensionless parameters
\begin{equation}
\mathcal{U}=\frac{A_{\mathrm{cap}}\sqrt{\alpha/D_t}} {\mathcal{L}}\quad 
\chgrev{\mathrm{and}\quad \mathcal{W}=\frac{\alpha \ell^2}{D_t}}
\label{eq:curlyU}
\end{equation}
where $A_{\mathrm{cap}}$ is the area of the capillary interface within the villus \chgrev{and $\ell$ is a lengthscale (that we compute) relevant to solute uptake under flow-limited conditions}.  We show how the transition from flow-limited to diffusion-limited transport, which occurs when $\mathrm{Da}^{-1}\sim 1\chgrev{+\mathcal{W}}$ when $\mathcal{U}\ll 1 $, instead occurs when $\mathrm{Da}^{-1}\sim \mathcal{U}$ when $\mathcal{U}\gg 1$.  We then extend our study to consider nonlinear uptake kinetics associated \chgrev{specifically} with oxygen metabolism, exploiting parameters that we determine from \textit{ex vivo} perfusion measurements, and examine the influence of flow on the partitioning of oxygen fluxes between placental tissue and fetal blood.  \chgrev{Our results suggest that oxygen uptake by terminal villous tissue has surprisingly limited impact on oxygen flux to the fetus.}

\section{Methods}

\begin{table*}[t!]
\vspace{-0.5em}
\centering
\caption{Reference parameter values used in the model (see Table~\ref{tab:oxygen-kinetics-comparison} for comparison of O$_2$ kinetics in different tissues and Table~\ref{tab:geometric_parameters} for villous geometric quantities). 
}
\begin{tabular}{l c c c}
\toprule
Parameter      &  Units                     & Value                      & Reference \\
\midrule  \addlinespace
$\Delta P$     &  $\units{Pa}$               & $\sim 10\,-\,10^2\quad\;\,$ &  \cite{erlich2018physical}
\tabularnewline
$\eta$         &  $\units{Pa \cdot s}$       & \quad\!$2\times10^{-3}$ &   \cite{erlich2018physical}
\tabularnewline
$B_{\chgrev{\text{, O$_2$}}}$   &         & \!\!\!$1.4\times10^{2} $ & \cite{pearce2016image}
\tabularnewline
$q_\text{max\chgrev{, O$_2$}}$ & $\units{mol/(m^3\cdot s)}$ & $\sim \chgrev{10^{-3}} - 10^{-1}$  & 
\chgrev{\cite{Schneider00,Schneider15,FrySecomb13}}
\tabularnewline
\chgrev{$q_\text{max{, glucose/fructose}}$} & \chgrev{$\units{mol/(m^3\cdot s)}$} & \chgrev{$\sim 10^{-3} - 10^{-2}$}  & 
\chgrev{\cite{Wang99,Nilsson74}}
\tabularnewline
$c_{50\chgrev{\text{,\,O$_2$}}}$  & $\units{mol/(m^3)}$              & $\sim 10^{-3} - 10^{-1}$    & 
\chgrev{\cite{FrySecomb13,Buerk78}}
\tabularnewline
$c_\text{mat\chgrev{, O$_2$}}$   & $\units{mol/(m^3)}$  & \quad\;\,$0.7\!\times\!10^{-1}$\; ($\approx 50\units{mmHg}$) &  \cite{pearce2016image}
\tabularnewline
\chgrev{$c_\text{mat, glucose}$}   & \chgrev{$\units{mol/(m^3)}$}  & \;\;\chgrev{$\sim 1 - 10$} & \cite{Hwang15,Wang99}
\tabularnewline
\chgrev{$c_\text{mat, fructose}$}  & \chgrev{$\units{mol/(m^3)}$}  & \chgrev{$\sim 10^{-2} - 10^{-1}$} &  \cite{Hwang15,Yap11}
\tabularnewline
$D_\text{t\chgrev{, O$_2$}}$  & $\units{m^2/s}$            & $1.7 \times 10^{-9}$          &  \cite{pearce2016image}
\tabularnewline
\chgrev{$D_\text{t, glucose}$}  & \chgrev{$\units{m^2/s}$}  & \chgrev{$\sim 10^{-12} - 10^{-11}$}  & 
\chgrev{\cite{erlich2018physical}}
\tabularnewline
\chgrev{$D_\text{t, fructose}$}  & \chgrev{$\units{m^2/s}$}  & \chgrev{$\sim 10^{-13} - 10^{-12}$}  & 
\chgrev{\cite{erlich2018physical}}
\tabularnewline\addlinespace
\chgrev{$\mathcal{U}_\text{O$_2$}$} &     & \:\chgrev{$\sim 0.01 - 0.1$} & \tabularnewline\addlinespace
\chgrev{$\mathcal{U}_\text{glucose}$} &   & \!\!\!\chgrev{$\sim \chgrevtwo{0.01} - 1$} & \tabularnewline\addlinespace
\chgrev{$\mathcal{U}_\text{fructose}$} &  & \quad\;\chgrev{$\sim 1 - 10$} & \tabularnewline\addlinespace
\bottomrule
\end{tabular}
\label{tab:model_params}
\end{table*}

\subsection{A mathematical model for feto-placental transport}

We summarise the computational model briefly here, providing technical details in Appendix~\ref{secAppendix:full-coupled-problem}. We model steady-state solute transport in an intravillous feto-placental capillary network as an advection-diffusion-uptake problem, extending existing models \cite{erlich2018physical,pearce2016image,mayo2016computational} to account for tissue metabolism. 

\chgrev{Three-dimensional images of villous microvasculature and the accompanying syncytiotrophoblastic shell (see Figs.~\ref{fig1:features-of-transport} and \ref{fig2:metabolism-CFD-vs-regression}, insets) were segmented and meshed from stained confocal microscopy of four specimens taken from two different peripherial lobules of a normal human placenta at term, as reported previously \cite{erlich2018physical,mayo2016computational,mayo2016three}.}  For each \chgrev{villous} specimen, the images reveal the spatial domain $\Omega_\mathrm{b}$ occupied by the capillary network.  This is bounded by an inlet surface $\Gamma_\mathrm{in}$, an outlet surface $\Gamma_\mathrm{out}$ and the capillary endothelium $\Gamma_{\mathrm{cap}}$ (see Fig.~\ref{fig:boundary-conditions-schematic}A in Appendix~\ref{secAppendix:full-coupled-problem}). 
The network is embedded in villous tissue, with exterior surface $\Gamma_{\mathrm{vil}}$, representing the interface with maternal blood.  A fixed solute concentration $c_\text{mat}$ is specified at  $\Gamma_\text{vil}$. Using a Newtonian (Stokes flow) approximation, our model simulates the flow of fetal blood entering through $\Gamma_\mathrm{in}$ and leaving via $\Gamma_\mathrm{out}$, driven by a pressure difference $\Delta P$ imposed between the inlet and outlet.  In the fetal capillaries, solutes are advected by blood flow and undergo diffusion. \chgrev{In the villous tissue that forms the bulk between the capillary surface $\Gamma_\mathrm{cap}$ and villous surface $\Gamma_\mathrm{vil}$,} we assume there is no flow and the solute concentration $c$ is assumed to obey a diffusion-uptake problem
$D_\text{t}\nabla^{2}c = q(c)$,
where $D_\text{t}$ is the solute diffusivity in tissue and $q(c)$ is the tissue solute metabolic rate.  In the present study we consider first a generic solute (such as a dilute suspension of polystyrene \chgrev{nano-}particles or other environmental pollutants \cite{Grafmaller_etal13,Stefanie_etal15}) characterised by first-order kinetics 
\begin{equation}
D_\text{t}\nabla^{2}c = \alpha\,c\,, 
\label{eq:Helmholtz}
\end{equation}
where $\alpha$ is the rate of \chgrev{solute uptake by tissue}\chgrevtwo{, assumed uniform}. 
We then model oxygen metabolism, using nonlinear Michaelis--Menten 
kinetics~\cite{Secomb_etal04,Bassingthwaighte97} 
\begin{equation}
D_\text{t}\nabla^{2}c = q_\text{max}\,\frac{c}{c_{50} + c}\,,
\label{eq:nonlin-kinetics}
\end{equation}
where $q_\text{max}$ is the maximum rate of oxygen metabolism and $c_{50}$ is the concentration at which the metabolic rate reaches 50\% of its maximum (Table~\ref{tab:model_params} summarises parameter values from the literature). For $c\ll c_{50}$, \eqref{eq:nonlin-kinetics} approaches \eqref{eq:Helmholtz} with $\alpha =  q_\text{max} / c_{50}$.  \chgrev{In keeping with prior physiological literature \citep{Buerk78}, \eqref{eq:nonlin-kinetics} can also be approximated (more empirically) using $\alpha \approx q_{\max}/c_{\mathrm{mat}}$.  Note that under linear kinetics, using (\ref{eq:Helmholtz}), transport depends on $\alpha$ and $D_t$ in the combination $\alpha/D_t$, as reflected in the parameters $\mathcal{U}$ and $\mathcal{W}$ in (\ref{eq:curlyU}).}


 Our computational model for 3D flow and transport was implemented in \texttt{COMSOL Multiphysics}$^\text{\textregistered}$ 5.3a, as described in \chgrev{\cite{erlich2018physical}}. A specific challenge of the modeling of transport is the emergence of boundary layers within the tissue when the uptake rate is high \chgrev{(corresponding to $\mathcal{U}\gg 1$)}, which required \chg{a particularly fine mesh resolution near the villous surface}. The meshes of the villous domain used in Figs.~\ref{fig1:features-of-transport}--\ref{fig3:region-diagram} below had approximately 20 million tetrahedral elements. Due to the weaker upatake in Fig.~\ref{fig4:nonlinear-metabolism}, a less detailed mesh was required (315k tetrahedral elements). In the latter case, a mesh convergence test revealed a change in the solute uptake of at most 2\% upon increasing the number of tetrahedral elements from 315k to 4 million. 
 
 In addition to full advection-diffusion-uptake computations, we employed a set of simulations of transport by diffusion and uptake alone, satisfying (\ref{eq:Helmholtz}) subject to simplified boundary conditions appropriate to flow-limited and diffusion-limited transport (described in Appendix~\ref{secAppendix:asymptotic-transport-regimes}).

\subsection{\emph{Ex vivo} measurement of placental oxygen metabolism}

In order to inform models of oxygen transport, we conducted experiments in order to estimate values of $q$ and $c_{50}$ \chgrev{for use in (\ref{eq:nonlin-kinetics})}.  All tissues were acquired from \chgrev{two full-}term human placentas delivered at St Mary’s Hospital, Manchester, UK, with appropriate informed written consent and ethical approval (15/NW/0829). \emph{Ex vivo} dual perfusion was established in an isolated lobule, as described previously \cite{Brownbill_etal11,Nye2018}. Briefly, each placental lobule was perfused via a peristaltic pump at an inflow rate of $14 \units{ml/min}$, oxygen concentration of 21\% (volume percent in air) delivered via a single cannula from the maternal side, and at $6 \units{ml/min}$, 0\% O$_2$ from the fetal side. Oxygen in tissue was recorded using a needle-type optical oxygen sensor (PyroScience FireStingO2 OXF500PT; Aachen, Germany) with an outer diameter of $500\units{{\mu}m}$ and the diameter of the tip of $230\units{{\mu}m}$. The optical sensors were 2-point calibrated as per the manufacturer’s instructions.

To record the tissue oxygen metabolic rate, the lobule was perfused until the oxygen reading reached a steady value. The oxygen drop-off curve (Fig.~\ref{fig:ex-vivo-metabolism-curve}A in Appendix~\ref{secAppendix:experiments} below) was recorded after cessation of both maternal and fetal inflows, while the needle-type oxygen probe was held at a fixed position approximately $8\units{mm}$ below the decidual surface. The measured oxygen decay rate was fitted to a nonlinear Michaelis--Menten law to estimate parameter values; details are provided in Appendix~\ref{secAppendix:experiments}.

\section{Results}

\subsection{Transport with linear uptake kinetics}
\label{sec:linear-kinetics-results}

\begin{figure*}[!t] 
\centering
\includegraphics[width=0.8\textwidth]{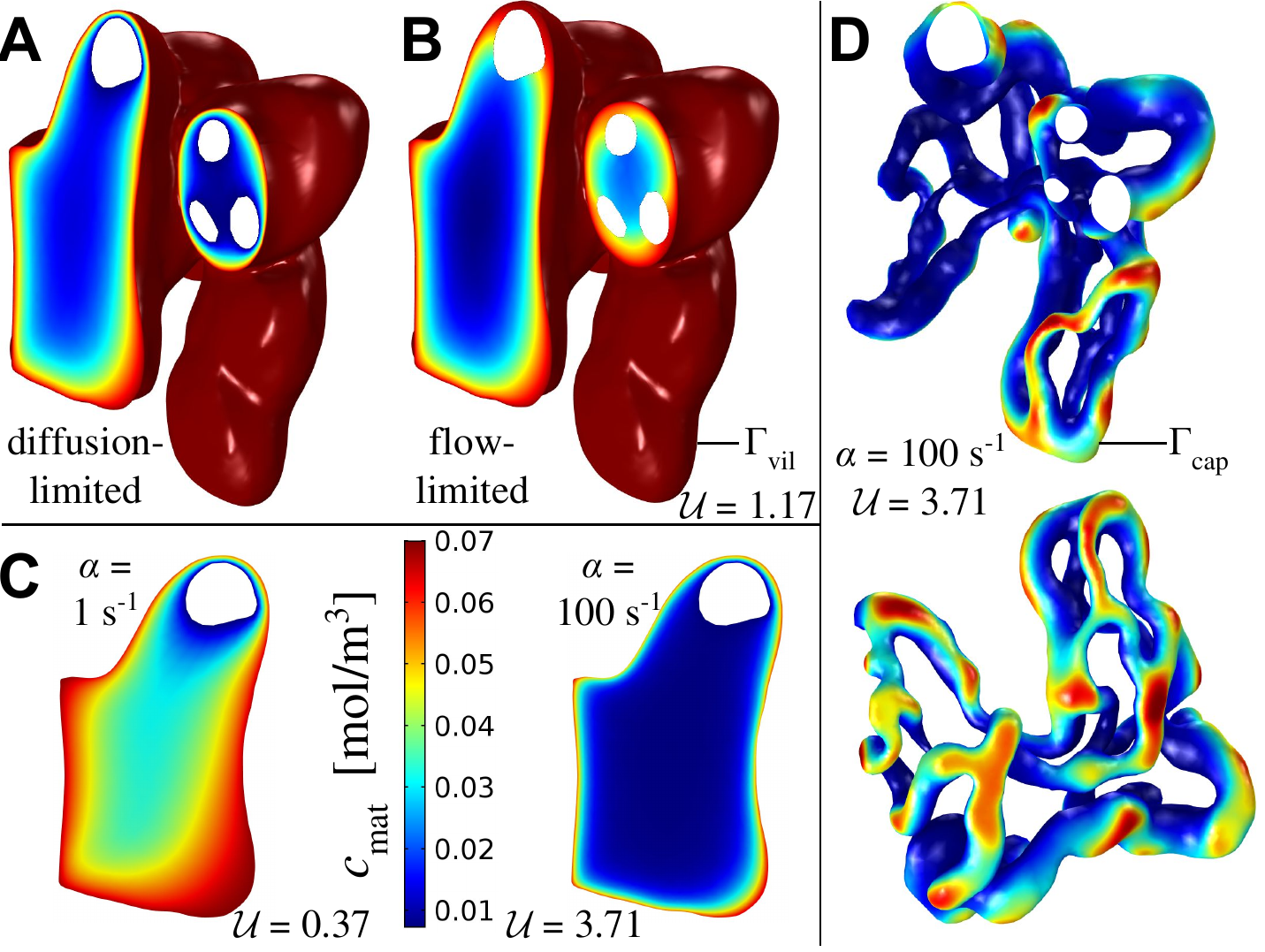}
\caption{\label{fig1:features-of-transport} Features of transport \chgrev{of a generic solute} in a terminal villus, \chgrev{assuming first-order kinetics}. \textbf{A, B}: Concentration fields in tissue are shown in a slice through specimen 3 under diffusion-limited and flow-limited conditions at metabolic rate $\alpha=10\units{s^{-1}}$. 
The villous surface is fully oxygenated due to the condition $c=c_\mathrm{mat}$ on $\Gamma_\mathrm{vil}$ (shown in dark red).  Vessel \chgrev{cross-sections} appear as white inclusions.  
\textbf{A} In the \chgrev{extreme} diffusion-limited case, $c=0$ at the capillary surface $\Gamma_\mathrm{cap}$. \textbf{B} In the \chgrev{extreme} flow-limited case, $\boldsymbol{n}\cdot \nabla \boldsymbol{c}=0$ at $\Gamma_\mathrm{cap}$. \textbf{C} Concentration slices (over part of the same surface shown in A, B) for the diffusion-limited case, with uptake rate $\alpha$ ranging over two orders of magnitude.  Concentration boundary layers form at the villous surface $\Gamma_\mathrm{vil}$ as $\alpha$ increases. 
\textbf{D} `Hotspots' emerge with increasing metabolism: \chgrev{only} where $\Gamma_\mathrm{cap}$ and $\Gamma_\mathrm{vil}$ are in close proximity \chgrev{can solute penetrate to capillaries}. {The top figure shows the capillary surface $\Gamma_\mathrm{cap}$ of Specimen 3 in the same spatial orientation as panels A \& B, with \chgrev{vessel cross-sections} shown in white. Colours show the concentration at $\Gamma_\mathrm{cap}$ for the extreme flow-limited case. 
The lower panel shows a different projection of the same simulation.} 
}
\end{figure*}

We first consider a model of linear uptake kinetics \chgrev{for a generic solute, as described by} \eqref{eq:Helmholtz}. Predictions of the computational model are shown in Fig.~\ref{fig1:features-of-transport}.  To show the possible range of behaviour, we first show the extreme cases of diffusion-limited transport (with negligible solute concentration in the capillary, \chgrev{a limit addressed in two spatial dimensions in Ref.~\cite{gill2011modeling}}) and flow-limited transport (when the solute concentration in the capillary equilibrates with the surrounding tissue, \chgrev{so that there is neglibible flux across the capillary surface}); these simplified limits are described in more detail in Appendix \ref{secAppendix:asymptotic-transport-regimes}.   In both cases, the solute concentration falls with distance from the villous surface, but does so more rapidly under diffusion-limited conditions (Fig.~\ref{fig1:features-of-transport}A, B).  These concentration fields were computed assuming a moderate metabolic rate ($\alpha=10\,\mathrm{s}^{-1}$, for which $\mathcal{U}\approx1.17$).  The impact of changing $\alpha$ is demonstrated in Fig.~\ref{fig1:features-of-transport}C, which shows how, under diffusion-limited conditions, 
concentration gradients become steeper as the uptake rate increases.  For sufficiently large $\alpha$ \chgrev{(\hbox{i.e.} $\mathcal{U}\gg 1$)}, 
most transport is reduced to a thin boundary layer (of thickness $\sqrt{D_{\mathrm{t}}/\alpha}$) near the villous surface, significantly reducing the solute flux reaching more internal capillaries.
This is illustrated in Fig.~\ref{fig1:features-of-transport}D, which shows the solute concentration at the capillary surface $\Gamma_\mathrm{cap}$ \chg{for specimen 3} in the flow-limited regime (the same scalebar applies as in Fig.~\ref{fig1:features-of-transport}C). 
The solute concentration on $\Gamma_{\mathrm{cap}}$ (and therefore the flux  across the capillary surface) is highly heterogeneous in this example. As tissue metabolism increases, localized regions of concentration (`hotspots') become more pronounced and solute transport becomes increasingly focused at a few regions at which the distance between the capillary and villous surfaces is locally minimal. 

\begin{figure*}[!tbp]
\centering
\includegraphics[width=1\textwidth]{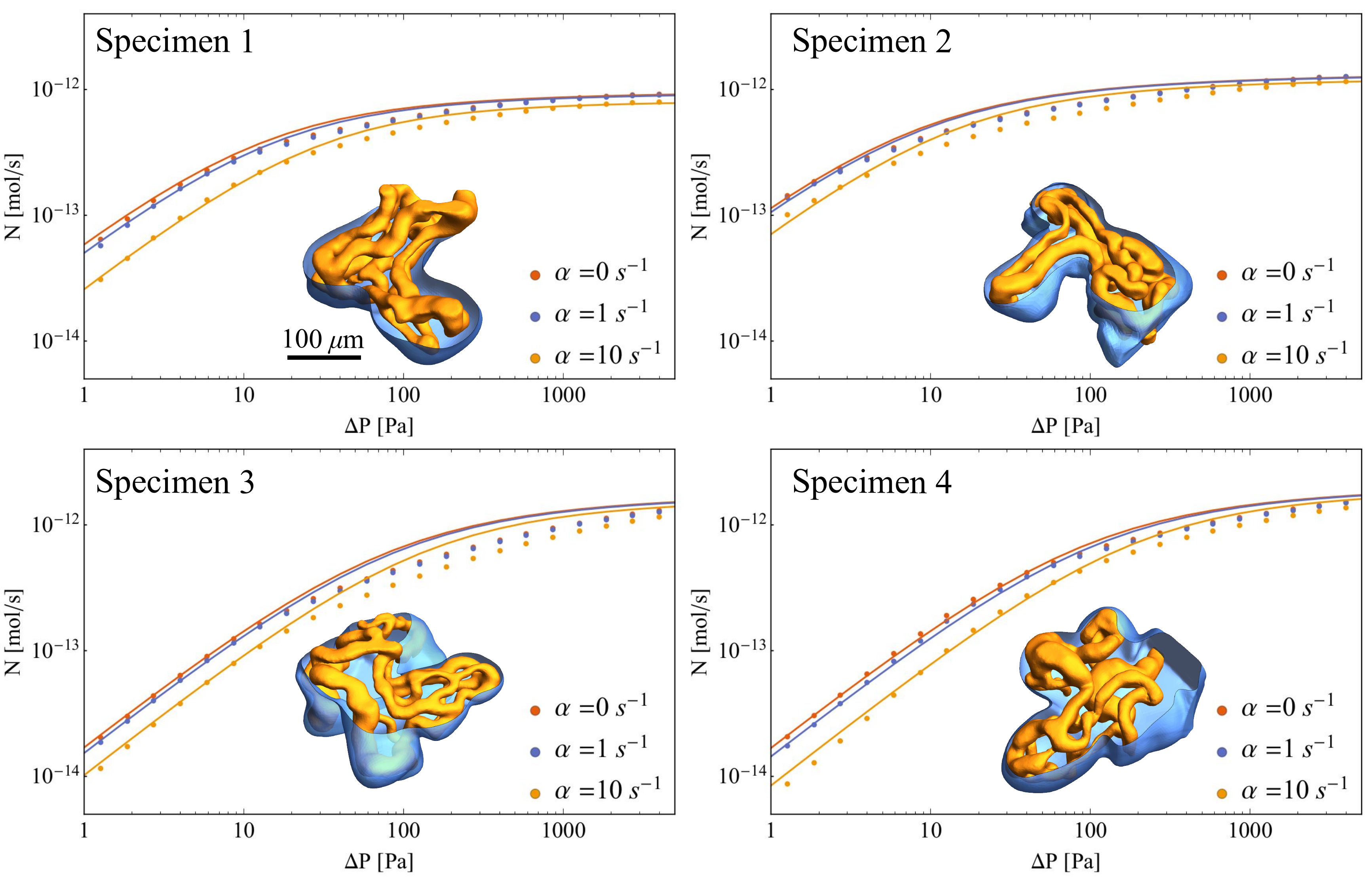}
\caption{\label{fig2:metabolism-CFD-vs-regression}Symbols show predictions of computational simulations of the full advection-diffusion-uptake problem (Appendix~\ref{secAppendix:full-coupled-problem}); curves show predictions of the \chgrev{simple} regression equation \eqref{eq:regression-equation-with-uptake}. The four panels show results for the four vasculatures used in \cite{erlich2018physical}; capillaries \chg{and \chgrevtwo{the} villous surface are illustrated by insets for  each case in orange and blue, respectively}. Each panel shows the net uptake $N$ as a function of the inlet-outlet pressure drop $\Delta P$, for no solute uptake ($\alpha=0$, identical with Fig. 2 in \cite{erlich2018physical}) and increasing uptake ($\alpha=1,\,10\units{s^{-1}}$\chgrev{; $D_\text{t}=1.7 \times 10^{-9} \units{m^2/s}$}).  
}
\end{figure*}

The symbols in Fig.~\ref{fig2:metabolism-CFD-vs-regression} show computational predictions of the solute flux $N$ entering four fetal capillary networks versus the pressure drop $\Delta P$ driving flow through the network, obtained using the full advection-diffusion-uptake model for the four specimens investigated.  In each case, $N$ rises approximately linearly with small $\Delta P$ (under flow-limited conditions) before saturating at large $\Delta P$ (under diffusion-limited conditions).  In the absence of uptake, we can use Eq.~\eqref{eq:regression-classical} to describe the flux/pressure-drop relationship: $N\approx N_\mathrm{max}/\mathrm{Da}\propto \Delta P$ when $\mathrm{Da}^{-1}\ll 1$, and $N\approx N_\mathrm{max}$ when $\mathrm{Da}^{-1}\gg 1$, where $N_\mathrm{max}$ is specific to each villus (see Appendix \ref{secAppendix:asymptotic-transport-regimes}). The symbols in Fig.~\ref{fig2:metabolism-CFD-vs-regression} also show that the impact of increasing the uptake parameter $\alpha$ is to reduce $N$ by an amount that diminishes slightly as $\Delta P$ increases. \chgrev{Overall, the change in metabolic uptake from $\alpha = 0$ to $\alpha = 1 \units{s^{-1}}$ causes a relative decrease in solute net uptake $N$ of at most 16\%, considering all pressure drops across all four specimens.}

To extend these predictions beyond the specific cases studied, it is helpful to approximate the relationship between $N$ and $\Delta P$ using a regression equation that incorporates relevant geometric parameters as well as the effect of metabolism.  \chgrev{(Our regression strategy is to identify simple algebraic expressions that capture key relationships with reasonable accuracy, rather than unwieldy but more precise formulae.)}  To generalise Eq.~\eqref{eq:regression-classical}, we first focus on how metabolism affects the \chgrev{extreme} flow-limited and diffusion-limited transport fluxes $N_\mathrm{FL}$ and $N_\mathrm{DL}$.  Computing these cases independently for each specimen, we determine the metabolic \chg{dimensionless} scale functions $G(\alpha)$ and $F(\alpha)$ \chg{that vary between 0 and 1} (Appendix \ref{secAppendix:metabolic-scale-functions}, Fig.~\ref{fig:metabolism-scale-function}) for which 
\begin{equation}
N_{\mathrm{FL}}=N_{\mathrm{max}}\mathrm{Da}^{-1}G(\alpha),\qquad N_{\mathrm{DL}}=N_{\mathrm{max}}F(\alpha).
\label{eq:FL-and-DL-fluxes}
\end{equation}
A simple algebraic approximation for \chgrev{generic} solute uptake under  linear kinetics across \chgrev{both} flow-limited and diffusion-limited transport regimes is then provided by \chgrev{constructing} the 
harmonic mean of 
$N_{\mathrm{FL}}$ and $N_{\mathrm{DL}}$ as 
\begin{align}
N&=\frac{1}{1/N_{\mathrm{DL}}\,+\,1/N_{\mathrm{FL}}\chgrev{\,+\,(\mathrm{Da}_F^{1/3}/N_{\max})}} \nonumber\\
&=\frac{N_{\mathrm{max}}}{1/F(\alpha)\,+\,\mathrm{Da}/G(\alpha)\chgrev{\,+\,\mathrm{Da_F}^{1/3}}}.
\label{eq:regression-equation-with-uptake}
\end{align}
This result provides an approximation for the net flux through any villus, requiring only a small set of computations of Eq.~(\ref{eq:Helmholtz}) under different boundary conditions, from which $\mathcal{L}$ (and hence $N_{\max}$), $F$ and $G$ can be determined. 
Fig.~\ref{fig2:metabolism-CFD-vs-regression} shows that  Eq.~\eqref{eq:regression-equation-with-uptake} provides a reasonable approximation of numerical solutions of the full advection-diffusion-uptake problem.  \chgrev{(The boundary layer correction $\mathrm{Da_F}^{1/3}$ in \eqref{eq:regression-equation-with-uptake} is not essential to this argument but it improves the accuracy of the approximation at intermediate $\mathrm{Da}$).} \chgrev{The largest relative error between computational results and equation \chgrevtwo{\eqref{eq:regression-equation-with-uptake}} across all specimens and all pressure drops is $33\%$ and occurs in Specimen 3.}


\begin{figure}[!tbp]
\centering
\includegraphics[width=0.5\textwidth]{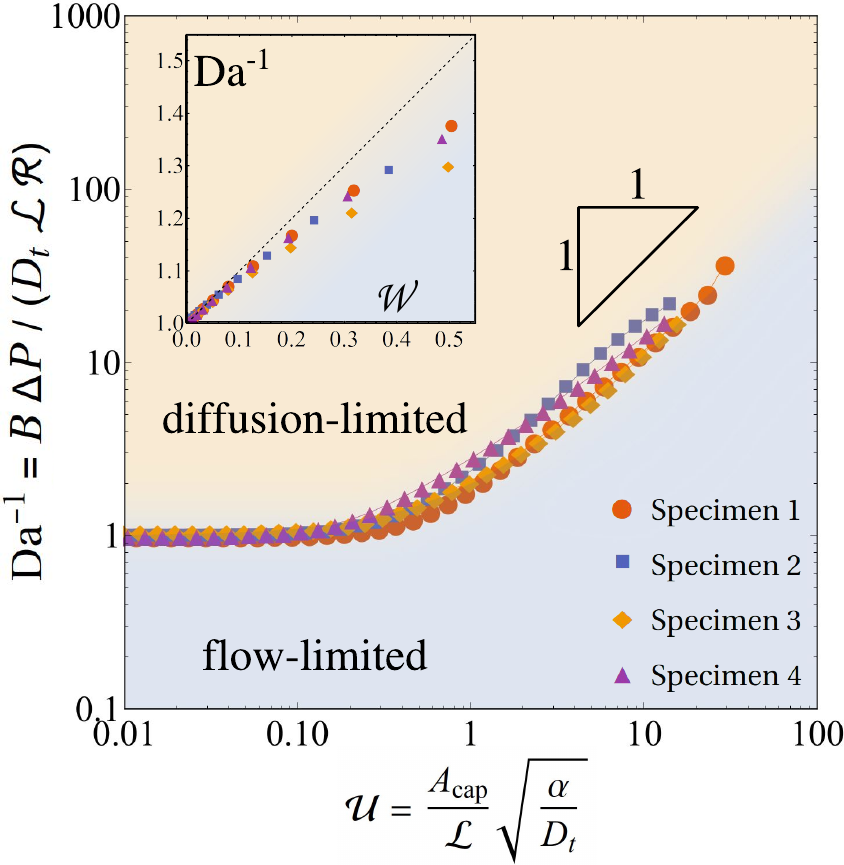}
\caption{\label{fig3:region-diagram}Flow-limited and diffusion-limited regimes in the presence of metabolism. The curves show for all four specimens the location where flow-limited and diffusion-limited regimes balance (\chgrev{which we define by} $\mathrm{Da}^{-1}=F/G$), as a function of the non-dimensional uptake parameter $\mathcal{U}$. For weak metabolism ($\mathcal{U}\ll 1$), $\mathrm{Da}^{-1}\approx 1$; for strong uptake ($\mathcal{U}\gg 1$), $\mathrm{Da}^{-1}\approx \mathcal{U}$. The boundary between flow-limited and diffusion-limited regimes should be understood as a smooth transition, indicating where contributions of both regimes are of comparable strength. The curves for all four specimens collapse appreciably. \chgrev{The inset illustrates how, when uptake is weak, the boundary between flow- and diffusion-limited uptake is approximated more precisely by $\mathrm{Da}^{-1}\approx 1+\mathcal{W}$ (dashed line), highlighting $\mathcal{W}$ as a significant dimensionless measure of uptake in this regime.} \chgrev{Geometric parameters are reported in Table~\ref{tab:geometric_parameters} below.}
}
\end{figure}

The regression equation (\ref{eq:regression-equation-with-uptake}) reveals the factors defining the transition between flow- and diffusion-limited transport.  Specifically, this arises when the two \chgrev{primary} terms in the denominator are of comparable magnitude, i.e. when $\mathrm{Da}^{-1}\approx F(\alpha)/G(\alpha)$.  This relationship is plotted for all four specimens in Fig.~\ref{fig3:region-diagram}.  The thresholds between the diffusion-limited and flow-limited regimes collapse onto a near-universal curve when $F/G$ is plotted not against $\alpha$ but instead against the nondimensional parameter $\mathcal{U}$ given in Eq.~\eqref{eq:curlyU}. This parameter emerges from an analysis of the large-$\alpha$ limit, described in Appendices~\ref{secAppendix:metabolic-scale-functions} \& \ref{secAppendix:WKB-approximation}.  

\chgrev{For weak metabolism ($\mathcal{U}\ll 1$), the solute distribution resembles that in the no-uptake limit discussed in \cite{erlich2018physical}, for which $F$ and $G$ are both close to unity.  Uptake in this case takes place across the whole volume of the villus tissue; this limit is examined further in Appendix~\ref{secAppendix:weak-metabolism} and is relevant to oxygen transport, as explained below.  
For solutes that are taken up strongly by villous tissues, delivery to fetal blood takes place via hotspots, i.e. local minima in the distance between $\Gamma_{\mathrm{cap}}$ and $\Gamma_{\mathrm{vil}}$ that are sufficiently small to penetrate the solute boundary layer adjacent to $\Gamma_{\mathrm{vil}}$ \chgrevtwo{(as illustrated in Fig.~\ref{fig1:features-of-transport}D)}.  We show in Appendices~\ref{secAppendix:metabolic-scale-functions} \& \ref{secAppendix:WKB-approximation} how common features between flow- and diffusion-limited transport explain the scaling relationship $F/G\approx \mathcal{U}$ when $\mathcal{U}\gg 1$.  Both functions decay exponentially fast as $\mathcal{U}$ increases (with metabolism becoming the dominant barrier to delivery to the fetus), but $G$ falls off faster than $F$ with increasing uptake (Figure~\ref{fig:metabolism-scale-function}), lowering the $N$ versus $\Delta P$ curve more at low flow-rates than at high flow rates, and hence promoting flow-limited transport relative to diffusion-limited transport when uptake is sufficiently strong.  

In addition to the lengthscale $\mathcal{L}$ (identified in our previous study \cite{erlich2018physical}), the capillary surface area $A_\mathrm{cap}$ becomes an important geometric determinant when solute uptake is strong (Appendix~\ref{secAppendix:WKB-approximation}).  In contrast, when solute uptake is weak (Appendix~\ref{secAppendix:weak-metabolism}) we identify an independent geometric quanity (the lengthscale $\ell$, appearing in the parameter $\mathcal{W}$ in (\ref{eq:curlyU})) which captures the weak-uptake approximation $\mathrm{Da}\approx 1+\mathcal{W}$ (see \chgrevtwo{the} inset to Fig.~\ref{fig3:region-diagram}). $\ell^2$ is determined by solving Poisson's equation over the villous volume, and is a normalised measure of the solute reduction by uptake through the tissue, communicated to the internal capillaries by diffusion.}    


\chgrev{In summary, we have shown how uptake of a generic solute in villus tissue under first-order kinetics reduces the rate of delivery of solute to the fetus (Figure~\ref{fig2:metabolism-CFD-vs-regression}).  We have quantified this reduction under flow-limited and diffusion-limited conditions, showing a greater impact in the former case and implying that stonger fetal flows are needed to achieve maximal delivery in the presence of uptake (Figure~\ref{fig3:region-diagram}). By using appropriate dimensionless parameters ($\mathrm{Da}$, $\mathcal{U}$ and $\mathcal{W}$), we have identified relationships that are independent of the details of individual villous geometries.  We now specialise our study to consider the important case of oxygen transport and uptake.}

\subsection{Nonlinear oxygen metabolism}
\label{sec:nonlin-kinetics-results}

\begin{figure*}[!tbp]
\centering
\includegraphics[width=1\textwidth]{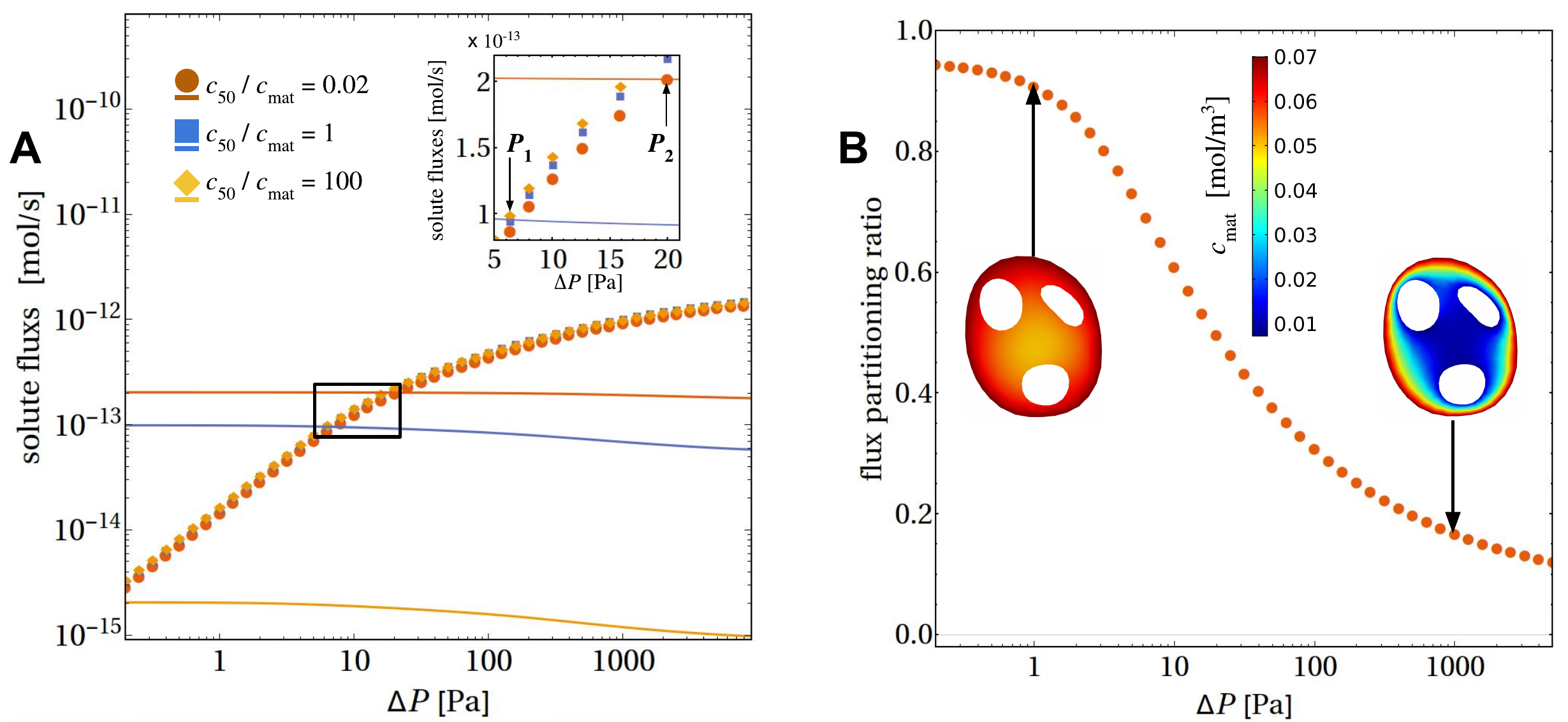}
\caption{\label{fig4:nonlinear-metabolism}The effect of non-linear uptake kinetics on solute transport. \textbf{A} The  \chg{coloured symbols} show the solute uptake $N$ \chg{which is delivered to the fetus}, plotted  against the inlet-outlet pressure drop $\Delta P$, for Specimen 3. \chg{For comparison, the solid lines with matched colours show the solute metabolised by villous tissue for the same specimen}. All curves are plotted for the same maximum rate of oxygen metabolism $q_{\mathrm{max}}=0.1\,\mathrm{mol}/(\mathrm{m}^{3}\cdot\mathrm{s})$, which we identify as a physiological value in metabolising tissue (Table \ref{tab:model_params}). The different curves show changes in the parameter $c_{50}$, spanning from predominantly zeroth-order \chg{(red symbols and curve)} to predominantly first-order \chg{(yellow symbols and curve)} uptake kinetics.   Intersections in the inset, marked $P_1$ and $P_2$, show where the metabolised oxygen flux balances the flux delivered to fetal blood.  
\textbf{B} The flux \chgrev{partitioning} ratio \chg{$N_\t{tissue}/N_\t{total}$} \chgrev{of the} solute \chgrev{flux} metabolised by the villous tissue \chgrev{to} the total flux of solute entering the terminal villus \chgrev{(for \chgrev{$c_{50}/c_\text{mat}=0.02$}; see }
Appendix~\ref{secAppendix:full-coupled-problem}). The flux partitioning ratio depends strongly on the flow regime. 
}
\end{figure*}

To establish the effect of metabolism on oxygen transfer in a realistic physiological context, we consider a Michaelis--Menten reaction-kinetics model for solute uptake in villous tissue, Eq.~\eqref{eq:nonlin-kinetics}. The model parameters are given in Table \ref{tab:model_params}. In particular, the Michaelis--Menten parameters $q_\mathrm{max}$ and $c_{50}$ are informed by a novel \textit{ex vivo} oxygen measurement study discussed in detail in Appendix \ref{secAppendix:experiments}. The parameter values emerging from this experimental study are compared with literature values in other metabolising tissues in Table \ref{tab:oxygen-kinetics-comparison}.
\chg{
The \emph{ex vivo} measured maximal rate of placental tissue metabolism $q_\t{max}$ agrees with the lower end of metabolic activity reported in other tissues (Table \ref{tab:oxygen-kinetics-comparison}, with the brain and cardiac tissue being more metabolically active). However \chgrevtwo{\emph{ex vivo} estimates suggest high variability in the `half-maximal metabolic rate' concentration}
$c_{50}$\chgrevtwo{, which} appears \chgrevtwo{larger} 
in the human placenta compared to other tissues \chgrevtwo{(Table~\ref{tab:oxygen-kinetics-comparison})}, indicating that a first-order kinetics approximation could be appropriate in less oxygenated regions of the intervillous space.}

Fig\chgrevtwo{ure}~\ref{fig4:nonlinear-metabolism}A shows the predicted solute uptake $N$ of the fetal capillary network versus the inlet-outlet pressure drop $\Delta P$ using the Specimen 3 geometry \chg{(coloured symbols)}. We consider three cases: $c_{50} \ll c_\mathrm{mat}$ \chg{(red symbols)}, when we expect to recover zeroth-order kinetics; $c_{50} = c_\mathrm{mat}$ \chg{(blue symbols)}, when the nonlinearity in the Michaelis--Menten approximation should be most apparent; and $c_{50} \gg c_\mathrm{mat}$ \chg{(yellow symbols)}, when we expect to recover first-order kinetics. Although there is uncertainty in the value of $c_{50}$ (Table \ref{tab:model_params}), the case $c_{50}/c_\mathrm{mat}=0.02$ (red symbols in Fig.~\ref{fig4:nonlinear-metabolism}A) lies within the range of physiological values of metabolising tissues (Table \ref{tab:model_params}). As the inset shows, the variation in $c_{50}$ across four orders of magnitude (spanning a transition from predominantly zeroth-order to predominantly first-order kinetics) reveals only modest variation in net oxygen delivery to fetal blood, \chg{affecting no more than 12\% change at an intermediate drop of $\Delta P=10\units{Pa}$}. \chg{This is in contrast to the solute flux metabolised by the villous tissue (solid lines, matching colours), \chgrevtwo{for} which the increase of $c_\mathrm{50}/c_\mathrm{mat}$ from 0.02 to 100 leads to a decrease in  the metabolised flux by two orders of magnitude.} We thus conclude that oxygen delivery to fetal blood in physiological conditions is not strongly affected by variation in $c_{50}$.  
\chgrev{Correspondingly, for a physiologically relevant oxygen metabolism rate $\alpha \sim 10^{-1} \units{s^{-1}}$ \cite{Buerk78}, we estimate the non-dimensional transport parameters \eqref{eq:curlyU} $\mathcal{W} \sim 10^{-2}$ and  $\mathcal{U}\sim 10^{-1}$ (see Tables~\ref{tab:model_params} and \ref{tab:geometric_parameters}), placing oxygen transport in the weak metabolism regime.
}

\begin{table*}[t!]
\centering
\caption{
Characteristic kinetic parameters for oxygen metabolism in different tissues. Oxygen solubility is taken to be $1.35\times10^{-3}\units{mol/(m^3\cdot mmHg)}$ \cite{BeardBassingthwaighte01}, 
tissue density is $\approx 10^{3} \units{kg/m^3}$, and the molar volumetric content of oxygen (at 37\,\textcelsius) is taken equal to $\approx 40 \units{{\mu}mol/(ml O_2)}$.
}
\begin{tabular}{l l l c}
\toprule
\multirow{2}{*}{Tissue}  &  $q_\text{max}$,  &  $c_{50}$,  &  \multirow{2}{*}{Reference}
\tabularnewline 
  &  $\units{mol/(m^3\cdot s)}$\, (\!$\units{ml/(kg\cdot min)}$)\quad\   
  &  $\units{mol/m^3}$\, (\!$\units{mmHg}$) & \\
\midrule  \addlinespace
Brain (mouse, \emph{in vivo}) & $\sim 10^{-2}$  &  & \cite{Devor_etal16} 
\tabularnewline
\chgrev{Brain (rat, \emph{ex vivo})} & $\sim 10^{-1}$ ($\approx\!140$) & $\sim 10^{-3}$ ($\approx\!0.8$) & \chgrev{\cite{Buerk78}} 
\tabularnewline
Brain & $\sim 10^{-1}$ ($140$)  & $\sim 10^{-3}$ ($1$) & \cite{FrySecomb13,Secomb_etal04} 
\tabularnewline
\chgrev{Liver (rat, \emph{ex vivo})} & $\sim 10^{-1}$ ($\approx\!88$) & $\sim 10^{-3}$ ($\approx\!2.2$) & \chgrev{\cite{Buerk78}} 
\tabularnewline
Tumour & $\sim 10^{-2}$ ($15$)  & $\sim 10^{-3}$ ($1$) & \cite{FrySecomb13,Secomb_etal04} 
\tabularnewline
Cardiac parenchyma & $\sim 10^{-2} - 10^{-1}$  & $\sim 10^{-4}$ ($\approx\!0.05$) & \cite{BeardBassingthwaighte01} 
\tabularnewline
Placenta (human, \emph{ex vivo}) & $\sim 10^{-3} - 10^{-2}$ ($2 - 11$)  & & \cite{Schneider00,Schneider15} 
\tabularnewline
Placenta (\emph{ex vivo}, this study)\quad\ & $\sim 10^{-2}$ ($\sim 10$) & $\sim 10^{-2} - 10^{-1}$ ($\sim 10 - 10^2$) &
\tabularnewline
\bottomrule
\end{tabular}
\label{tab:oxygen-kinetics-comparison}
\end{table*}

Fig.~\ref{fig4:nonlinear-metabolism}B shows how the partitioning of fluxes between \chg{villous} tissue and fetal blood depends on the flow regime. The ratio of fluxes \chg{$N_\t{tissue}/N_\t{total}$} is defined as the amount of \chgrevtwo{solute} metabolised by the villous tissue divided by the total flux that enters into the terminal villous from the maternal circulation through the villous surface \chg{(see Appendix~\ref{secAppendix:full-coupled-problem}, Eq.~\eqref{eq:relative_tissue_consumption})}.
In the flow-limited regime (\chgrevtwo{with} very low values of $\Delta P$), almost all of the solute entering the terminal villus is metabolised by the villous tissue. Conversely, in the diffusion-limited regime (\chgrevtwo{with} high values of $\Delta P$), a small fraction of total flux is metabolised, \chgrevtwo{and} a larger fraction of the solute enters the fetal capillary. This highlights how the flux is partitioned differently depending on the flow regime. 
\chg{
For a physiological range of terminal pressure drops of $\sim 10-100 \units{Pa}$, the model predicts relative oxygen consumption by terminal villous tissue of approximately 30\% to 60\% of the total oxygen supply, which is comparable to the upper range of $22-54$\% reported as the relative oxygen consumption rate by the human placenta \emph{ex vivo} and \emph{in vivo}~\cite{Carter00,Schneider00}.
}


\chgrev{In summary, parameter estimates from measurements of dynamic oxygen uptake rate in placental tissue and our computational model together suggest that the rate at which oxygen is metabolised by a terminal villus is substantially smaller than the maximum (diffusion-limited) transfer rate of the villus, and also sufficiently small for oxygen to penetrate throughout the villous tissue.  However under strongly flow-limited conditions, our model predicts that villous tissue can absorb a substantial proportion of the oxygen supplied from the maternal circulation.}

\section{Discussion}

Computational models of physiological function are important both in developing fundamental scientific understanding and in advancing medical therapies.  Like many organs, the placenta has a complex multiscale organisation that challenges current methodologies.  We present here a set of results \chgrevtwo{for} terminal villi, the primary functional exchange units of the fetal circulation, in which we use 3D simulations to derive simplified expressions of solute transport that can be readily integrated within larger-scale models of solute transport.  Despite high variability among the four villus samples investigated, we have shown that a handful of geometric statistics are sufficient to characterise transport of solutes that are taken up by the villous tissue itself.  It is hoped that these results will guide future studies of microvascular anatomy so that function can be assessed more readily from 3D structural datasets. 

\chgrev{Our strategy in the present study has been to explore a broad range of parameters, to illustrate possible outcomes for a variety of solutes (see Table~\ref{tab:model_params}), before focusing attention on oxygen.  One benefit of this approach is that the simplifications emerging for extreme parameter values shed light on underlying physical mechanisms, which then help us understand the more complex interactions that emerge under physiological conditions.}

In formulating approximations of solute exchange, we have sought to use dimensionless quantities that naturally characterise dominant physical balances.  In the absence of solute uptake in tissue, we showed previously \cite{pearce2016image, jensenchernyavsky2018, erlich2018physical} that the Damk{\"o}hler number (\ref{eq:da}) is useful in distinguishing flow-limited from diffusion-limited transport (with the transition between the two cases occurring when Da is of order unity).  Additional parameters emerge when solute metabolism is accounted for.  When solute uptake (assuming first-order kinetics) is sufficiently strong to induce solute boundary layers within villous tissue, the relevant uptake parameter is $\mathcal{U}$ (see (\ref{eq:curlyU})), as illustrated in Fig.~\ref{fig3:region-diagram}.  For more moderate uptake (a limit of relevance to oxygen), however, a further parameter $\mathcal{W}$ emerges, as shown in Appendix~\ref{secAppendix:weak-metabolism}.  Distinct geometric quantities appear in each parameter, reflecting the differing physical balances: $\mathcal{L}$ in Da measures a mean exchange area over exchange distance, as is appropriate to diffusion-limited transport; the total area of the capillary endothelium $A_\mathrm{cap}$ in $\mathcal{U}$ is relevant to transport under flow-limited conditions when fetal blood is exposed to a varying concentration field over this surface; \chgrev{and the lengthscale $\ell$ appears in $\mathcal{W}$, reflecting uptake of solute throughout the bulk of the tissue.  Our study demonstrates how solutions of simple canonical partial differential equations (here, the 3D Laplace \& Poisson equations) in complex spatial domains can be used to extract these functionally significant geometric measures from imaging data.}  



Our linear and non-linear uptake models allow insight into how the transport of different solutes is affected by metabolism. A key finding of the linear uptake model for a generic solute 
is that sufficiently strong uptake can drive solute exchange towards the flow-limited regime~(Fig.~\ref{fig3:region-diagram}). \chgrev{This may be relevant for certain sugars: our estimate of $\mathcal{U}$ for fructose (in excess of unity, Table~\ref{tab:model_params}) suggests that metabolism can have a strong impact on its exchange.}
To test the effect of metabolism on oxygen transport, we implemented a nonlinear Michaelis--Menten uptake model and parametrised it with physiological literature values from different metabolising tissues, including our own experiments on placental tissue using an oxygen probe (Fig.~\ref{fig:ex-vivo-metabolism-curve}).   Our simulations predict that oxygen transport to fetal blood is only modestly affected by metabolism (Fig.~\ref{fig4:nonlinear-metabolism}\chgrev{A}).  This implies that for physiological values, 
\chgrev{zero-uptake predictions (such as \citep{erlich2018physical}) provide viable leading-order estimates of oxygen delivery to the fetus, allowing us to determine the impact of oxygen uptake by tissue as linear corrections (see (\ref{eq:N_DL}) and (\ref{eq:N_FL}) \chgrevtwo{below}).  Nevertheless,}
the metabolic flux 
\chgrev{is much larger in comparison to}
the flux delivered to fetal blood at low fetal flow rates \chgrev{(more precisely, when $\mathrm{Da}^{-1}\lesssim \mathcal{W}\ll 1$, see Appendix~\ref{secAppendix:weak-metabolism})}, and \emph{vice versa} at high flow rates (Fig.~\ref{fig4:nonlinear-metabolism}\chgrev{B}).

The present model rests on numerous assumptions.  We demonstrated previously \cite{erlich2018physical} that non-Newtonian effects of fetal blood flow can be neglected in a first approximation, although numerous features of oxygen transport by red blood cells and dynamic hematocrit distribution in complex networks require further assessment \cite{Hellums_etal95}.  Clearly there will be value in performing additional studies in a wider sample of villus networks, in order to test the robustness of the present approximations and to consider the impact of structural \chg{and metabolic} abnormalities that may arise in disease.  Future studies should also address the maternal flow exterior to the villus surface, to test the assumption that the source of solute is uniformly distributed \chgrevtwo{and to identify appropriate lengthscales that determine transport} (see also \cite{serov2015analytical,serov2015optimal}).  Given that there is spatial heterogeneity across the whole placenta, the kinetics may switch from zeroth order to first order in different locations within the same organ. Active transport of some solutes by the syncytiotrophoblast is a further refinement that will be required to make robust predictions of placental function.  \chgrev{Prior studies of transport in other physiological systems \cite{sapoval1994, sapoval2002, grebenkov2005, grebenkov2006} suggest that there is value in using a mixed (Robin) boundary conditions on $\Gamma_{\mathrm{cap}}$ to explore states between the extremes considered in Appendices \ref{secAppendix:asymptotic-transport-regimes} and \ref{secAppendix:metabolic-scale-functions}.} Finally, our study considers uptake only in \chgrev{terminal villous} branches, and does not account for solute metabolism by other placental tissues, which will influence overall delivery to the fetus.


{In summary, this study offers an integrated approach to characterise the transport of solutes, such as oxygen, that are metabolised by tissue with complex embedded microvasculature. A robust algebraic relationship (\ref{eq:regression-equation-with-uptake}) provides a computationally efficient tool to upscale micro-structural features to the organ-scale function of the human placenta\chgrevtwo{; this approach should be adaptable to} other physiological systems with complex vasculature. Although our realistic image-based model offers a general insight into relative contributions of villous tissue metabolism, diffusive capacity and feto-capillary flow, more data are needed to further quantify the identified transport determinants in healthy and abnormal placentas.
}

\begin{acknowledgements}
{\small
The authors are grateful to Romina Plitman Mayo (\chgrevtwo{Tel Aviv University, Israel}) for sharing the villous geometries, and we thank Edward D. Johnstone (University of Manchester\chgrevtwo{, UK}) and Rohan M. Lewis (University of Southampton\chgrevtwo{, UK}) for helpful discussions.\\ 

\chgrevtwo{
\textbf{Funding}: This work was supported by  MRC research grant MR/N011538/1.

\textbf{Competing interests}: The authors declare that they have no competing interests.

\textbf{Author contributions}:  AE, OEJ and ILC took part in mathematical model design; GN, PB and ILC conceived experimental model; AE performed numerical simulations; GN conducted \emph{ex vivo} experiments; AE, OEJ and ILC performed asymptotic and data analysis; AE, OEJ and ILC interpreted the results and prepared the manuscript. All authors read and approved the final manuscript.

\textbf{Data accessibility}: All data needed to evaluate the results and conclusions are present in the paper. The associated datasets and codes can be accessed via the Figshare repository (\url{http://doi.org/10.6084/m9.figshare.7718462}). Additional data related to this study may be available from the authors upon request.
}}
\end{acknowledgements}

\appendix

\section{\label{secAppendix:full-coupled-problem}The computational model}

\begin{figure}[!tbp] 
\centering
\includegraphics[width=0.5\textwidth]{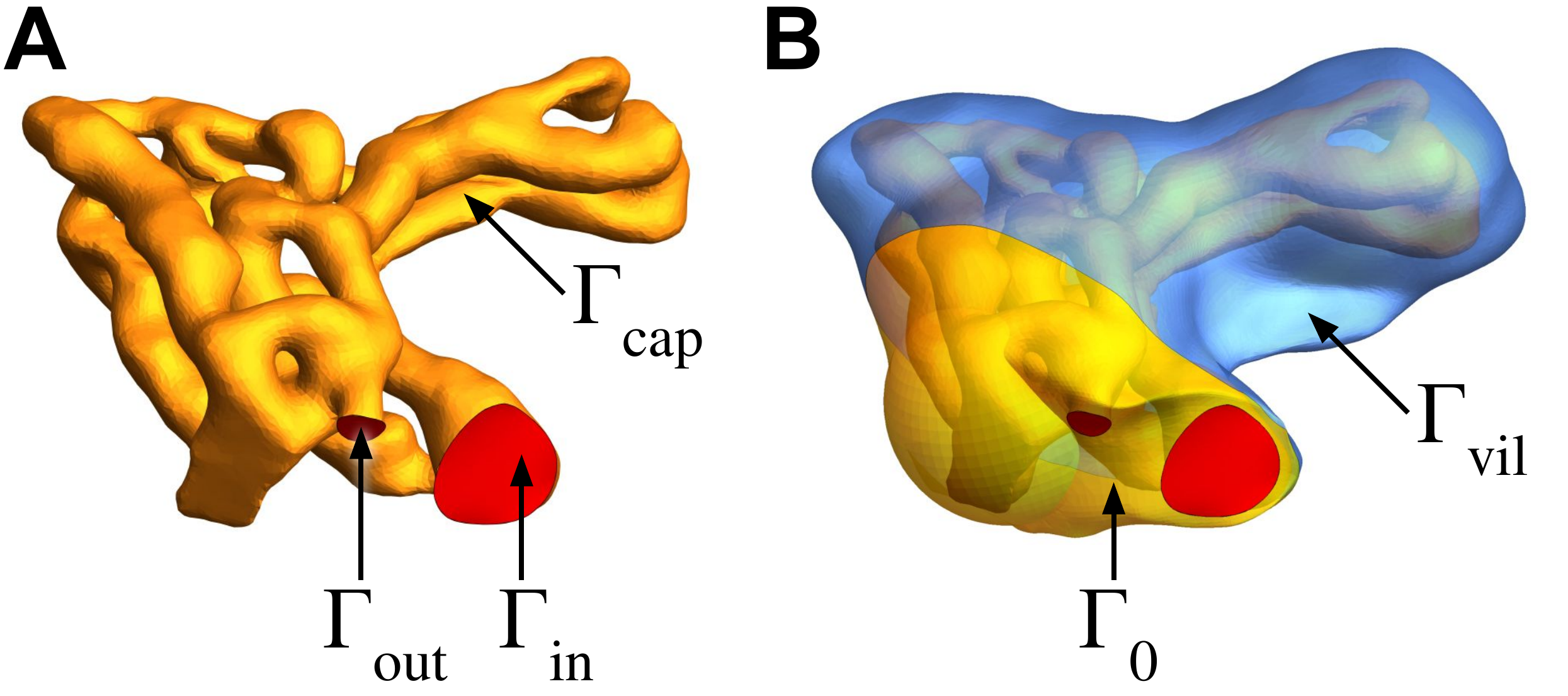}
\caption{Computational domain of specimen 1, and boundary surfaces $\Gamma$. \textbf{A} The domain occupied by blood vessels $\Omega_\mathrm{b}$ is bounded by the inlet and outlet surfaces $\Gamma_\mathrm{in}$ and $\Gamma_\mathrm{out}$ (red) and the capillary surface $\Gamma_\mathrm{cap}$ (yellow). 
\textbf{B} The domain occupied by villous tissue $\Omega_\mathrm{t}$ is bounded by the capillary surface $\Gamma_\mathrm{cap}$, the no-flux surface $\Gamma_0$, and the villous surface $\Gamma_\mathrm{vil}$ (blue). 
}\label{fig:boundary-conditions-schematic}
\end{figure}

We model fetal blood flow within a fetal capillary network using the Stokes equations, which are solved over the domain $\Omega_\mathrm{b}$ occupied by blood vessels: 
\begin{equation}
\eta\nabla^{2}\boldsymbol{u}=\nabla p,\qquad\nabla\cdot\boldsymbol{u}=0,\qquad \boldsymbol{x}\in \Omega_\mathrm{b}.
\label{eq:basic-Stokes-PDE}
\end{equation}
Here $\mathbf{x}$ is a spatial coordinate, $\boldsymbol{u}(\mathbf{x})$ is the fluid velocity field, $p(\mathbf{x})$ the fluid pressure and $\eta$ the dynamic viscosity of fetal blood, which is treated as Newtonian in our 3D simulations.  We take $\eta = 2\times 10^{-3} \units{Pa\!\cdot\!s}$ (Table~\ref{tab:model_params}), which is appropriate for blood with 48\% hematocrit in a 20$\units{\mu{m}}$ vessel (see 
\cite{erlich2018physical}). 
The capillary domain $\Omega_\text{b}$ is bounded by the inlet surface $\Gamma_\text{in}$, the outlet surface $\Gamma_\text{out}$, and the capillary surface $\Gamma_\text{cap}$, see Fig.~\ref{fig:boundary-conditions-schematic}A. \chgrev{The tissue domain} $\Omega_\mathrm{t}$ is bounded internally by the capillary surface $\Gamma_\mathrm{cap}$ and externally by the no-flux surface $\Gamma_0$ and the villous surface $\Gamma_\mathrm{vil}$, see Fig.~\ref{fig:boundary-conditions-schematic}B. 
Blood enters through the inlet surface $\Gamma_\text{in}$ and leaves via $\Gamma_\text{out}$, driven by a pressure difference $\Delta P$ imposed between inlet and outlet. A no-slip condition is imposed on the interior of $\Gamma_\text{cap}$. The boundary conditions on the flow are therefore
\begin{align}
p & =\Delta P\quad\text{on}\quad\Gamma_\mathrm{in},\label{eq:inletPressure}\\
p & =0\quad\text{on}\quad\Gamma_\mathrm{out},\label{eq:outletPressure}\\
\boldsymbol{u} & =0\quad\text{on}\quad\Gamma_\mathrm{cap}.\label{eq:noSlip}
\end{align}
The volume flux is defined as 
\begin{equation}
Q=\iintop_{\Gamma_{\mathrm{out}}}\boldsymbol{n}\cdot\boldsymbol{u}\,\mathrm{d}A
\label{eq:volume-flux-generic}
\end{equation}
and the resistance of the network is \begin{equation}
\mathcal{R}=\Delta P/Q.
\label{eq:resistance-generic}
\end{equation}
Within $\Omega_\mathrm{b}$, the solute concentration $c(\mathbf{x})$ is assumed to obey the linear advection-diffusion equation
\begin{equation}
B\boldsymbol{u} \cdot \nabla c = D_\mathrm{p} \nabla^2 c,\qquad \boldsymbol{x}\in \Omega_\mathrm{b},
\label{eq:advection-diffusion}
\end{equation}
where $D_\text{p}$ is the solute diffusion coefficient in plasma.  The parameter $B=1$ for most solutes, but $B>1$ for species that bind to hemoglobin, modelling facilitated transport by red blood cells. For oxygen \chgrev{in fetal blood}, $B\approx 141$ \cite{serov2015analytical,pearce2016image}. Fetal blood is assumed to enter solute-free at the inlet $\Gamma_\mathrm{in}$ and zero diffusive solute flux is imposed at the outlet $\Gamma_\mathrm{out}$:
\begin{align}
c=0 & \quad\text{on}\quad\Gamma_\text{in},\label{eq:inletConcentration}\\
\boldsymbol{n}\cdot\nabla c=0 & \quad\text{on}\quad\Gamma_\text{out}.
\label{eq:outletConcentration}
\end{align}
The capillaries are surrounded by villous tissue, which occupies the domain $\Omega_\mathrm{t}$. Here the solute concentration is assumed to obey a diffusion-uptake equation
\begin{equation}
D_\mathrm{t} \nabla^2 c = q(c),\qquad \boldsymbol{x}\in \Omega_\mathrm{t},
    \label{eq:diffusion-uptake}
\end{equation}
where $D_t$ is the solute diffusion coefficient in tissue and \chg{$q(c)$ is the solute uptake rate, which is taken either to satisfy first-order kinetics \chgrev{for a generic solute}, $q(c) = \alpha c$ (Section~\ref{sec:linear-kinetics-results} of the Results), or to obey the non-linear Michaelis--Menten relationship \eqref{eq:nonlin-kinetics} \chgrev{for oxygen} (Section~\ref{sec:nonlin-kinetics-results}). Here} $\alpha$ is a rate of metabolic uptake; $\alpha = 0$ reproduces the case discussed in \cite{erlich2018physical}. 

\begin{figure*}[!tbp]
\centering
\includegraphics[width=\textwidth]{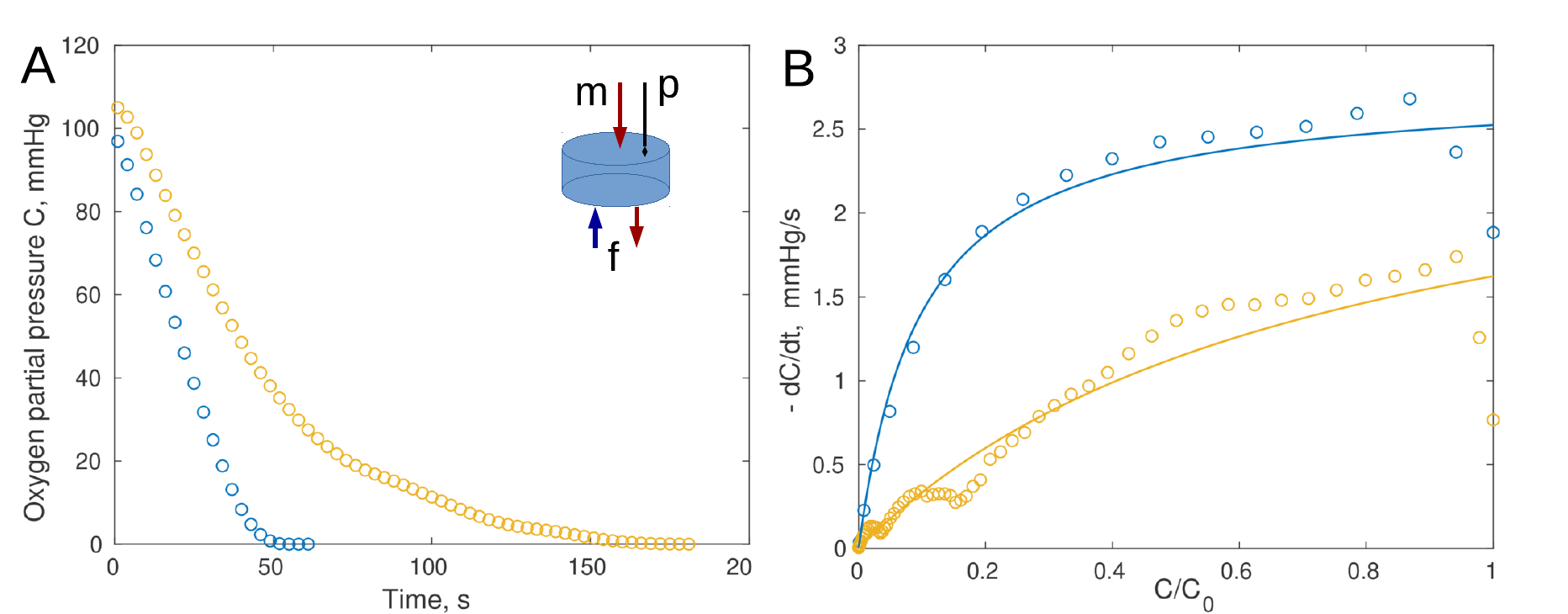}
\caption{Oxygen metabolism kinetics in the human placenta \emph{ex vivo}. \textbf{A} Tissue oxygen partial pressure decay after the cessation of the flow in \chgrev{two} dually-perfused placentas; inset shows the dual perfusion setup, with maternal (m) and fetal (f) cannulas and the optical sensor probe (p). \textbf{B} Fitted (\chgrev{Eq.~\eqref{eq:M-M-kinetics},} solid) \chgrev{vs.\ measured (circles)} oxygen metabolic rates. \chgrev{See Methods and Appendix~\ref{secAppendix:experiments} for more details.}}
\label{fig:ex-vivo-metabolism-curve}
\end{figure*}

The maternal solute concentration $c=c_\text{mat}$ is imposed on $\Gamma_\text{vil}$ and no solute flux is imposed at the intermediate region $\Gamma_0$ on the villous surface to avoid artificial sharp gradients (we assume that the geometric \chgrevtwo{domain} was sliced across $\Gamma_0$ from the larger network of a villous tree).  Thus
\begin{align}
\boldsymbol{n}\cdot\nabla c=0 & \quad\text{on}\quad\Gamma_0, \label{eq:diffout}\\
c=c_{\text{mat}} & \quad\text{on}\quad\Gamma_\text{vil}.\label{eq:maternalConcentration}
\end{align}
We couple the problems in $\Omega_\mathrm{b}$ and $\Omega_\mathrm{t}$ by imposing continuity of the concentration across $\Gamma_\mathrm{cap}$ 
as well as matching the diffusive fluxes
\begin{equation}
D_{\mathrm{p}}\,\boldsymbol{n}\cdot\nabla c=D_{\mathrm{t}}\,\boldsymbol{n}\cdot\nabla c\quad\text{on}\quad\Gamma_{\mathrm{cap}}.
\label{eq:flux-matching}
\end{equation}
Once a solution $c$ to \eqref{eq:basic-Stokes-PDE}--\eqref{eq:flux-matching} has been obtained, we compute the net solute flux delivered to the fetus, defined as the integral of the advective flux over the outlet boundary
\begin{equation}
N=\iintop_{\Gamma_{\mathrm{out}}}B\,c\,\boldsymbol{n}\cdot\boldsymbol{u}\,\mathrm{d}A.
\label{eq:net-flux-generic}
\end{equation}
Given (\ref{eq:inletConcentration}) and (\ref{eq:diffout}), $N$ is equal to the sum of diffusive fluxes across inlet and capillary surface:
\begin{equation}
N=\iintop_{\Gamma_{\mathrm{cap}}}\boldsymbol{n}\cdot\nabla c\,\mathrm{d}A-\iintop_{\Gamma_{\mathrm{in}}}\boldsymbol{n}\cdot\nabla c\,\mathrm{d}A.
\end{equation}
\chg{The proportion of the solute flux metabolised by the villous tissue relative to the total flux supplied by the maternal blood to the villous surface is given by
\begin{equation}
\frac{N_\t{tissue}}{N_\t{total}} = 
\frac{ \iiint_{\Omega_\t{t}} q(c)\d{V} }{ \iint_{\Gamma_\t{vil}}\vec{n}\cdot\nabla c\,\d{A} }\,.
\label{eq:relative_tissue_consumption}
\end{equation}
}

\section{\label{secAppendix:asymptotic-transport-regimes}Asymptotic transport regimes 
under first-order kinetics}

The computation of $N$ requires the numerical solution of a complex boundary value problem \eqref{eq:basic-Stokes-PDE}--\eqref{eq:flux-matching} over multiple domains, $\Omega_\mathrm{b}$ and $\Omega_\mathrm{t}$. In order to simplify the problem and gain physical understanding, we consider the simpler asymptotic regimes of \chgrev{extreme} diffusion- and flow-limited transport, in which the computation can be restricted to $\Omega_\mathrm{t}$.   Concentration profiles in each regime 
are shown in Fig.~\ref{fig1:features-of-transport}A, and are determined as follows. 

In \chgrev{extreme} diffusion-limited transport, the flow is sufficiently rapid to impose a fixed concentration difference $\Delta c \chg{= c_\t{mat}}$ between the villous boundary $\Gamma_{\text{vil}}$ and the capillary boundary $\Gamma_{\text{cap}}$. Assuming linear kinetics \chgrev{for a generic solute}, we define the diffusion-limited boundary value problem as
\begin{equation}
\begin{aligned}
D_{\mathrm{t}}\nabla^{2}c & =\alpha c & \text{on}\quad & \Omega_{\mathrm{t}}\\
c & =0 & \text{on}\quad & \Gamma_{\mathrm{cap}}\\
c & =c_{\mathrm{mat}} & \text{on}\quad & \Gamma_{\mathrm{vil}}\\
\boldsymbol{n}\cdot\nabla c & =0 & \text{on}\quad & \Gamma_{0}.
\end{aligned}
\label{eq:DL-PDE-generic}
\end{equation}
The net solute flux to fetal blood can be evaluated as
\begin{equation}
N_{\mathrm{DL}}=D_\mathrm{t}\iintop_{\Gamma_{\mathrm{cap}}}\boldsymbol{n}\cdot\nabla c\,\mathrm{d}A.
\label{eq:DL-flux-generic}
\end{equation}
This is a function of the uptake rate $\alpha$ as well as the domain shape. In the case $\alpha=0$, we retrieve the maximum exchange capacity $N_\mathrm{max}\equiv N_\mathrm{DL}\vert_{\alpha=0}$ from \cite{erlich2018physical}.
The lengthscale $\mathcal{L}$ associated with a villus is defined as $\mathcal{L}=N_\mathrm{max}/(D_\mathrm{t} c_\mathrm{mat})$. 

\chgrev{Extreme} flow-limited transport arises when $\Delta P$ is sufficiently weak for solute to be fully saturated in fetal blood before it leaves the vessel network.  In this case $N$ is proportional to the (weak) flow rate through the outlet. The flow-limited problem differs from \eqref{eq:DL-PDE-generic} by the boundary condition on the capillary surface, which we approximate by assuming that the flux to the vessel is negligible to leading order, 
\begin{equation}
\begin{aligned}
D_{\mathrm{t}}\nabla^{2}c & =\alpha c & \text{on}\quad & \Omega_{\mathrm{t}}\\
\boldsymbol{n}\cdot\nabla c & =0 & \text{on}\quad & \Gamma_{\mathrm{cap}},\:\Gamma_{0}\\
c & =c_{\mathrm{mat}} & \text{on}\quad & \Gamma_{\mathrm{vil}}.
\end{aligned}
\label{eq:FL-PDE-generic}
\end{equation}
Within the capillary, the concentration profile is almost equilibrated, and swept away by weak flow at the outlet.  We model solute mixing by advection and diffusion within the capillary by assuming that
the advective flux at the outlet (\ref{eq:net-flux-generic}) is determined by a surface average of 
the concentration obtained as a solution to \eqref{eq:FL-PDE-generic}
\begin{equation}
N_{\mathrm{FL}}\approx B\,Q\,\langle c\rangle\,, \quad
\chg{
\langle c\rangle = \frac{1}{A_\t{cap}} \iint_{\Gamma_\t{cap}}\!\! c\,\d{A}\,,
}
\label{eq:flow-limited-flux-approximated}
\end{equation}
where $\langle c \rangle$ is the mean concentration over the capillary surface $\Gamma_\mathrm{cap}$ and $Q$ is the volume flux \eqref{eq:volume-flux-generic}.


\section{\label{secAppendix:experiments}\emph{Ex vivo} oxygen metabolism in the human placenta}

We estimate local oxygen metabolism parameters, assuming Michaelis--Menten kinetics \cite{Secomb_etal04,Bassingthwaighte97} for the concentration $\bar{c}$ measured by the optical sensor, volume-averaged over a region of \chgrev{size $\sim 50\units{{\mu}m}$} in the intervillous space.  \chgrev{Provided the diffusion timescale across the measured volume is smaller than the tissue metabolic timescale (see Tables~\ref{tab:model_params} \& \ref{tab:oxygen-kinetics-comparison})}, the local response to a cessation of flow is assumed to satisfy
\begin{equation}
-\Dfrac{\bar{c}}{t} = \phi_\text{t}\,\bar{q}_\text{max}\,\frac{\bar{c}}{\bar{c}_{50} + \bar{c}}\,,
\label{eq:M-M-kinetics}
\end{equation}
where $\bar{q}_\text{max}$ and $\bar{c}_{50}$ are volume-averaged effective kinetic parameters, and $\phi_\text{t}$ is the villous volume fraction. 
 
We use nonlinear least squares fitting via the Levenberg--Marquardt algorithm implemented in \texttt{nlinfit} function of \texttt{MathWorks MATLAB}$^\text{\textregistered}$ R2018a to estimate the parameter values of \eqref{eq:M-M-kinetics} from experimental data (Fig.~\ref{fig:ex-vivo-metabolism-curve}). The fitted values of the volume-averaged parameter $\phi_\text{t}\bar{q}_\text{max}$ (mean $\pm$ SE) are $2.77 \pm 0.11 \units{mmHg/s}$ (Subject 1) and $2.8 \pm 0.3 \units{mmHg/s}$ (Subject 2), and the values for $\bar{c}_{50}$ vary from $9.3 \pm 1.6 \units{mmHg}$ to $78 \pm 16 \units{mmHg}$. Assuming $\phi_\text{t} \approx 0.5$ \cite{serov2015optimal} and oxygen solubility of $\approx 1.35 \times 10^{-3} \units{mol \cdot m^{-3} \cdot mmHg^{-1}}$ \cite{BeardBassingthwaighte01}, we obtain $q_\text{max} \sim 10^{-2} \units{mol \cdot s^{-1} \cdot m^{-3}}$ and
$c_{50} \sim 10 - 10^{2} \units{mmHg}$ (or, $\sim 10^{-2} - 10^{-1} \units{mol/m^3}$); see Table~\ref{tab:oxygen-kinetics-comparison} for a comparison of placental oxygen metabolism to other tissues. 

\chgrev{We emphasise that the reported values are from just two placentas and further studies are necessary to assess inter- and intra-placental variability.}

\section{\label{secAppendix:metabolic-scale-functions}Metabolic scale functions under 
first-order kinetics}

\begin{figure} 
\centering
\includegraphics[width=0.5\textwidth]{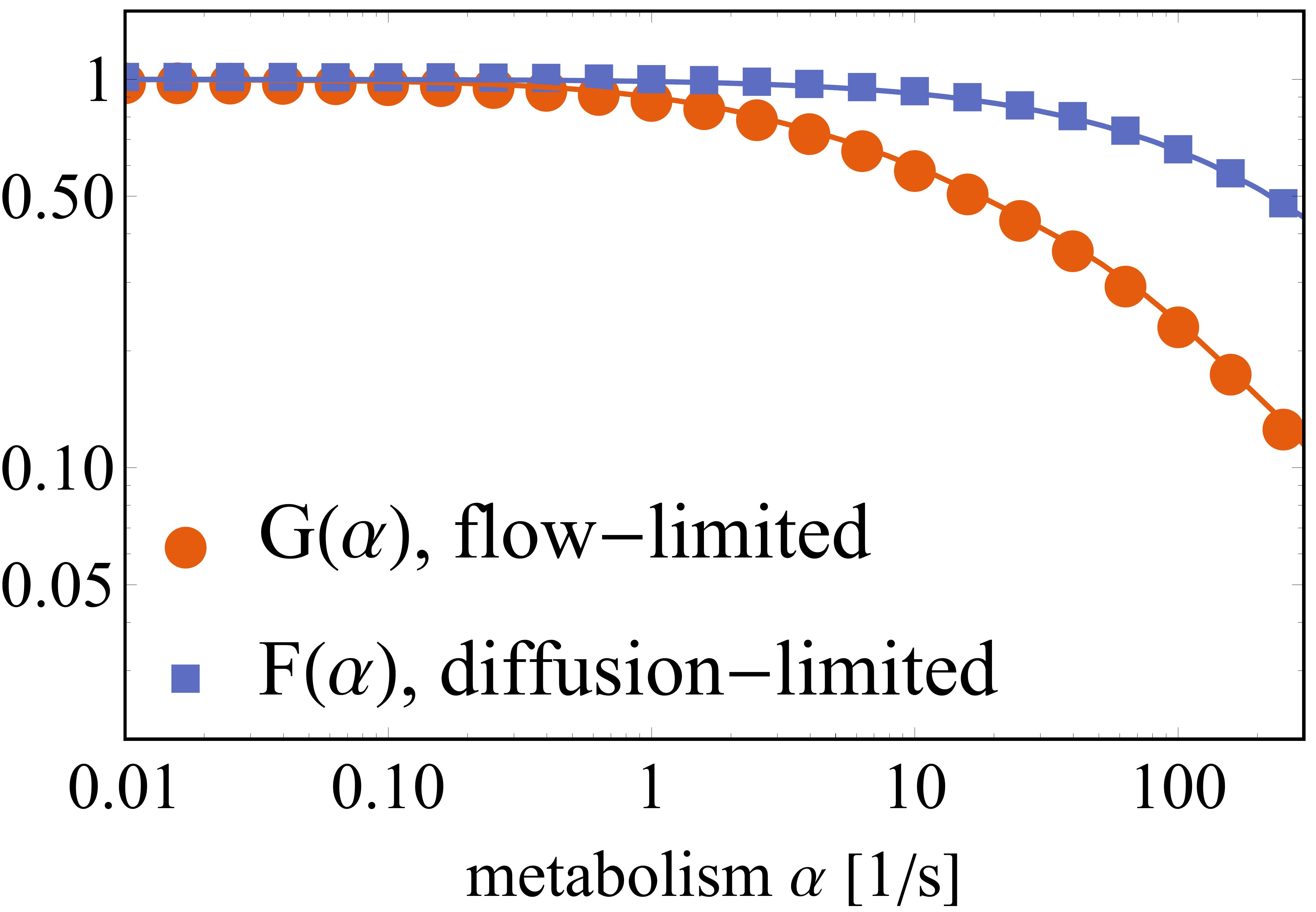}
\caption{\label{fig:metabolism-scale-function}Metabolic scale functions computed for specimen 3. By definition, $F=1$ and $G=1$ for the case of no metabolism, $\alpha = 0$. 
The points result from finite element calculations and the lines connecting the points are meant as visual guides.}
\end{figure}

\chgrevtwo{Extending the framework of Appendix~\ref{secAppendix:asymptotic-transport-regimes}, we}
can recast the diffusion-limited flux \eqref{eq:DL-flux-generic} and the flow-limited flux \eqref{eq:flow-limited-flux-approximated} into a more convenient form by introducing the dimensionless metabolic scale functions $0<F(\alpha)\leq 1$ and $0<G(\alpha)\leq 1$ satisfying
\begin{align}
N_{\mathrm{DL}}= & N_{\mathrm{max}}F(\alpha), \label{eq:DL-flux-with-F(alpha)}\\
N_{\mathrm{FL}}= & N_{\mathrm{max}}\mathrm{Da}^{-1}G(\alpha). \label{eq:FL-flux-with-G(alpha)}
\end{align}
These functions satisfy $F(0)=1$ and $G(0)=1$, retrieving the no-metabolism forms of flow-limited and diffusion-limited flux from \cite{erlich2018physical}.
The diffusion-limited metabolic scale function $F(\alpha)$ is defined in terms of an integral over the concentration $c$ which is the solution to the Helmholtz problem (\ref{eq:DL-PDE-generic}) 
\begin{equation}
F(\alpha)=
\frac{1}{c_{\mathrm{mat}}\mathcal{L}}
\iintop_{\Gamma_{\text{cap}}}\boldsymbol{n}\cdot\nabla c\,\mathrm{d}A\,.
\label{eq:DL-metabolic-scale-function-F}
\end{equation}
The flow-limited metabolic scale function $G(\alpha)$ is derived from the solution to the Helmholtz problem (\ref{eq:FL-PDE-generic}) as 
\begin{equation}
G(\alpha)=\frac{1}{c_\t{mat}\,A_\t{cap}}\iintop_{\Gamma_\t{cap}}c\,\mathrm{d}A\,.
\label{eq:FL-metabolic-scale-function-G}
\end{equation}
These functions are illustrated in Fig.~\ref{fig:metabolism-scale-function}, for one villus (Specimen 3).  Each function becomes exponentially small as the uptake parameter increases, because $\Gamma_\mathrm{cap}$ falls outside the concentration boundary layer adjacent to $\Gamma_\mathrm{vil}$.  The dependence of $F$ and $G$ on villous geometry is explored further below. 

\chgrev{To explain why $G$ falls off more rapidly \chgrevtwo{than} $F$ for large $\alpha$, consider that when uptake is strong the $c$ field is confined to a boundary layer of thickness $\sqrt{D_t/\alpha}\equiv 1/\theta$ adjacent to $\Gamma_{\mathrm{vil}}$, so that $F$ and $G$ will be dominated by localized regions (hotspots) in which $\Gamma_{\mathrm{cap}}$ is closest to $\Gamma_{\mathrm{vil}}$.  Suppose that in such a region the distance between the two surfaces can be represented by a paraboloid with mean radius of curvature $R$; then the region over which any uptake takes place is confined to distances $\sqrt{R/\theta}$ of the point of closest approach of the two surfaces, \hbox{i.e.} an area of size $R/\theta$.  $F$, being an area integral (\ref{eq:DL-metabolic-scale-function-F}) of $\mathbf{n}\cdot\nabla c$ where $c$ varies over a lengthscale $1/\theta$, will be proportional to $\theta(R/\theta)/\mathcal{L}=R/\mathcal{L}$; $G$, being an area integral (\ref{eq:FL-metabolic-scale-function-G}) of $c$, will be proportional to $(R/\theta)/A_{\mathrm{cap}}$.  As explained in further detail in Appendix~\ref{secAppendix:WKB-approximation} below, their ratio $F/G$ therefore scales like $\theta A_{\mathrm{cap}}/\mathcal{L}\equiv \mathcal{U}$ (see (\ref{eq:curlyU})) for $\mathcal{U}\gg 1$, implying $G\ll F$ in this limit.}



\section{\label{secAppendix:WKB-approximation}WKB approximation for strong metabolism}

We seek an approximations of (\ref{eq:DL-PDE-generic}) and (\ref{eq:FL-PDE-generic}) under strong uptake, assuming first-order kinetics.  Let $C=c/c_\mathrm{mat}$ and $\theta^2=\alpha/D_\t{t}$.  We consider the limit in which the boundary-layer thickness $1/\theta$ becomes smaller than the thickness of the villous tissue.  
%
%
We pose a WKB expansion, $C=\exp(\theta S_0 + S_1+S_2/\theta+\dots)$.  Then at successive orders, the diffusion-uptake equation gives
\begin{align}
\vert \nabla S_0 \vert^2 =1 \quad\mathrm{and}\quad
2 \nabla S_0 \cdot \nabla S_1 + \nabla^2 S_0 =0. \label{eq:tra}
\end{align}
We assume $\Gamma_{\mathrm{vil}}$ is parametrized by surface coordinates $\mathbf{s}\equiv (s_1,s_2)$ and a local normal coordinate $n$, oriented into the villous tissue.  Let $\hat{\mathbf{n}}(\mathbf{s})$ be the local unit normal and $\kappa(\mathbf{s}) = \nabla_s\cdot \hat{\mathbf{n}}$ the curvature, where $\nabla_s\equiv (\mathsf{I}-\hat{\mathbf{n}}\otimes \hat{\mathbf{n}})\cdot\nabla$.

The eikonal equation (\ref{eq:tra}a) is solved by $S_0=\pm n$, where $n$ measures distance along $\hat{\mathbf{n}}$.  The transport equation (\ref{eq:tra}b) becomes $2\hat{\mathbf{n}}\cdot \nabla S_1 + \kappa =0$, implying that $S_1=-\tfrac{1}{2} \kappa n+A(\mathbf{s})$ for some constant $A$.  Thus the  leading-order expression for $C$ can be written
\begin{equation}
C=A_+ \exp (n\theta-\tfrac{1}{2}\kappa n + \dots) + A_- \exp (-n\theta-\tfrac{1}{2}\kappa n + \dots).
\end{equation}
The boundary layer thickness $1/\theta$ must be significantly smaller than the wall radius of curvature $1/\kappa$ for this approximation to be valid.  Imposing $C=1$ on $\Gamma_{\mathrm{vil}}$ implies $A_++A_-=1$.  Ignoring complications arising from caustics, we assume $\Gamma_{\mathrm{cap}}$ lies at $n(\mathbf{s})=N(\mathbf{s})$.  The boundary condition on this surface determines $A_\pm$.  Let $\hat{\mathbf{m}}(\mathbf{s})$ denote the unit normal to $\Gamma_{\mathrm{cap}}$, pointing into the capillary.

Imposing $C=0$ on $\Gamma_{\mathrm{cap}}$ in the diffusion-limited case implies 
\begin{equation}
C=\frac{\exp(-n\theta-\tfrac{1}{2}\kappa n)}{1-\exp(-2N\theta)}
-\frac{\exp(n\theta-\tfrac{1}{2}\kappa n)}{\exp(2N\theta)-1}
\end{equation}
and so 
\begin{equation}
F\approx \frac{1}{\mathcal{L}} \int_{\Gamma_\mathrm{cap}} 2\, \hat{\mathbf{m}}\cdot\hat{\mathbf{n}}\,\theta e^{-N\theta-\kappa N/2}  \,\mathrm{d}A.
\label{eq:F}
\end{equation}
Imposing $C_n=0$ on $\Gamma_{\mathrm{cap}}$ in the flow-limited case implies 
\begin{equation}
C=\frac{\exp(n\theta-\tfrac{1}{2}\kappa n)}{1+\exp(2N\theta)}
-\frac{\exp(-n\theta-\tfrac{1}{2}\kappa n)}{1+\exp(-2N\theta)}
\end{equation}
and so 
\begin{equation}
G\approx \frac{1}{A_{\mathrm{cap}}} \int_{\Gamma_\mathrm{cap}} 2 e^{-N\theta-\kappa N/2}  \,\mathrm{d}A.
\label{eq:G}
\end{equation}

We can estimate the integrals in (\ref{eq:F}) and (\ref{eq:G}) by noting that the exponential will be dominated by the local minima of $N$.  Let such a point be at $\mathbf{s}=\mathbf{s}_0$, at which $\nabla_s N=\mathbf{0}$.  The present analysis addresses the case in which $N(\mathbf{s}_0)\gg 1/\theta$.  Let $\lambda_1>0$, $\lambda_2>0$ be the eigenvalues of the Jacobian $\nabla_s \otimes \nabla_s N(\mathbf{s}_0)$.   The local area element may be written as $\mathrm{d}A=\sqrt{1+\vert \nabla_s N \vert^2} \mathrm{d}s_1 \mathrm{d}s_2$, so that near $\mathbf{s}_0$ we have $\mathrm{d}A\approx \mathrm{d}s_1 \mathrm{d}s_2$.  We can rotate the coordinates locally so that the Jacobian is diagonal, and then $N\approx N(\mathbf{s}_0) + \tfrac{1}{2}(\lambda_1 (\mathbf{s}-\mathbf{s}_0)_1^2 + \lambda_1 (\mathbf{s}-\mathbf{s}_0)_2^2 )+\dots $.  Introducing scaled coordinates $u_i=(\theta \lambda_i)^{1/2} (\mathbf{s}-\mathbf{s}_0)_i$ for $i=1,2$, we obtain as a contribution to $G$
\begin{equation}
\frac{2 e^{-N(\mathbf{s}_0)\theta }}{A_{\mathrm{cap}}}\int_{-\infty}^{\infty}\int_{-\infty}^{\infty}   
\frac{e^{-\frac{1}{2} (u_1^2+u_2^2) }}{\theta \sqrt{ \lambda_1 \lambda_2}}\,\mathrm{d}u_1 \mathrm{d}u_2=\frac{4\pi R(\mathbf{s}_0)} {A_{\mathrm{cap}}\theta} e^{-N(\mathbf{s}_0)\theta}.
\label{eq:g}
\end{equation}
The factor $R(\mathbf{s}_0)\equiv 1/\sqrt{\lambda_1 \lambda_2}$ is a lengthscale associated with the mean radius of the `hotspot' at $\mathbf{s}_0$.  The prefactor in (\ref{eq:g}) is a dimensionless ratio of this length times the boundary-layer thickness $1/\theta$ to $A_{\mathrm{cap}}$, and is associated with the assumption that the flow absorbs solute from $\Gamma_{\mathrm{cap}}$ by (effectively) averaging over the capillary interface.  The exponential sensitivity to $N(\mathbf{s}_0) \theta$ in (\ref{eq:g}) shows that proximity of $\Gamma_{\mathrm{vil}}$ to $\Gamma_{\mathrm{cap}}$ is the predominant factor in determining overall exchange under flow-limited conditions with strong uptake.  We can expect $G$ to be dominated by contributions from a small number of such hotspots within a villus, each with its own value of $N(\mathbf{s}_0)$ and $R(\mathbf{s}_0)$.  

The corresponding contribution to $F$ will have $\hat{\mathbf{m}}\cdot\hat{\mathbf{n}}\vert_{\mathbf{s}_0}\approx 1$, giving
\begin{equation}
\frac{4\pi R(\mathbf{s}_0) }{\mathcal{L}} e^{-N(\mathbf{s}_0)\theta}.
\label{eq:f}
\end{equation}
In this case the net exchange capacity under diffusion-limited conditions, proportional to $\mathcal{L} F$, is determined instead by the hotspot radius $R(\mathbf{s}_0)$, becoming independent of the global measure $\mathcal{L}$.

In general, $F$ and $G$ will be determined by a sum of such contributions from the dominant 'hot-spots' in each case, but with \chgrevtwo{(from \eqref{eq:g} and \eqref{eq:f})}
\begin{equation}
\frac{F}{G}\approx  \frac{A_{\mathrm{cap}} \theta}{\mathcal{L}} \equiv \frac{A_\mathrm{cap}\sqrt{\alpha/D_\t{t}} }{\mathcal{L}}
\label{eq:F-G-ratio-large-alpha}
\end{equation}
for sufficiently large $\alpha$.   Thus we see the emergence of the dimensionless parameter $\mathcal{U}$ in (\ref{eq:curlyU}).  The relationship (\ref{eq:F-G-ratio-large-alpha}) is validated in Fig.~\ref{fig3:region-diagram}.

\begin{table}[t!]
\centering
\caption{Geometric parameters used in defining $\mathcal{U}$ and $\mathcal{W}$, computed assuming first-order kinetics (with $\alpha=q_{\max}/c_\mathrm{mat}$ and $f(C)=C$).  $\mathcal{L}$ was reported previously in \citep{erlich2018physical}. 
}
\begin{tabular}{p{6em} p{4em} p{4em} p{4em} p{4em}}
\toprule
Specimen & 1 & 2 & 3 & 4\tabularnewline
\midrule  \addlinespace
$\mathcal{L},\units{mm}$ & 8.2 & 11.4 & 15.4 & 17.9\tabularnewline
\addlinespace 
$A_\text{cap},\units{mm^2}$ & 0.125 & 0.0830 & 0.124 & 0.122\tabularnewline
\addlinespace 
$\mathcal{V},\units{mm^3}$ & 0.0016 & 0.0014 & 0.0021 & 0.0021\tabularnewline
\addlinespace 
$\ell_\text{DL},\units{mm}$ & 0.0056 & 0.0042 & 0.0050 & 0.0040\tabularnewline
\addlinespace
$\ell,\units{mm}$ & 0.0185 & 0.0128 & 0.0146 & 0.0181\tabularnewline
\addlinespace
$\mathcal{V}/(\mathcal{L}\ell^2)$ & 0.57 & 0.75 & 0.64 & 0.36\tabularnewline
\bottomrule
\end{tabular}
\label{tab:geometric_parameters}
\end{table}

\section{\label{secAppendix:weak-metabolism}Weak metabolism limit}

\chgrev{The case of weak tissue metabolism arises when the total rate of solute uptake by tissue, which we can estimate by $q_{\max}\mathcal{V}$ where $\mathcal{V}$ is the volume of villous tissue in $\Omega_\t{t}$, is smaller than the maximum possible flux reaching fetal blood $N_{\max}\equiv D_t c_{\mathrm{mat}}\mathcal{L}$.  Equivalently we require $q_{\max}/(D_\t{t}\,c_\t{mat})$ to be smaller than a quantity with dimensions of inverse area, that we estimate provisionally as $\mathcal{L}/ \mathcal{V}$ but which we now determine more precisely.  We can decouple the diffusive and uptake fluxes in \eqref{eq:diffusion-uptake} by expanding the solution of 
\begin{equation}
\nabla^2 C = \eps\,f(C),\quad \eps \equiv \frac{q_\t{max}}{D_\t{t}c_\t{mat}},\quad C \equiv \frac{c}{c_\t{mat}}\,,
\label{eq:weak-metabolism}
\end{equation}
formally in powers of $\varepsilon$, subject to appropriate boundary conditions, as $C \approx C^\0 + \eps\,C^\1+\dots$.
Here $f(C)$ is a non-negative dimensionless metabolic function (equal to $1$ for zeroth-order, {$C$ for first-order}, or $C/(C + c_{50}/c_\t{mat})$ for Michaelis--Menten kinetics).  While $C^\0$ is dimensionless, $C^\1$ has dimensions of length squared in this formulation and is expected to be negative, reflecting the reduction of the solute field by delivery to tissue.}

\chgrev{In diffusion-limited conditions with $C|_{\Gamma_\t{cap}} = 0$ and $C|_{\Gamma_\t{vil}} = 1$, we have, at leading order, $\nabla^2 C^\0 = 0$, $C^\0|_{\Gamma_\t{cap}} = 0$ and $C^\0|_{\Gamma_\t{vil}} = 1$, and thus the flux to fetal blood is $N_\t{DL}^\0 \equiv \int_{\Gamma_\t{cap}} \vec{n}\cdot\nabla C^\0 \d{A} = \mathcal{L}$. The following order gives $\nabla^2 C^\1 = f(C^\0)$, $C^\1|_{\Gamma_\t{cap}} = C^\1|_{\Gamma_\t{vil}} = 0$.  The reduction to the flux to fetal blood due to solute uptake in tissue reveals the effective area, which we write as the square of a lengthscale $\ell_{DL}$, {
\begin{equation}
\ell^2_{DL}= -\frac{1}{\mathcal{L}}\,\int_{\Gamma_\t{cap}}\! \vec{n}\cdot\nabla C^\1 \d{A},
\label{eq:weak-metabolism-DL}
\end{equation}}
which can be computed numerically for a given villous geometry (Table~\ref{tab:geometric_parameters}). In dimensional variables, the approximate total flux becomes {
\begin{equation}
N_\t{DL} \approx D_\t{t}\,c_\t{mat}\,\mathcal{L} - q_\t{max}\, \mathcal{L} \, \ell^2_{DL}, 
\label{eq:N_DL}
\end{equation}}
or equivalently (using \eqref{eq:DL-flux-with-F(alpha)}) {$F\approx 1-\varepsilon \ell^2_\mathrm{DL}$ for $\varepsilon \ell^2_\mathrm{DL}\ll 1$.}}

\chgrev{In the flow-limited case, {$\vec{n}\cdot \nabla C|_{\Gamma_\t{cap}} = 0$} and $C|_{\Gamma_\t{vil}} = 1$. Thus, at successive orders, \eqref{eq:weak-metabolism} gives $C^\0 \equiv 1$,
$N_\t{FL}^\0 \equiv \frac{1}{A_\t{cap}}\int_{\Gamma_\t{cap}} C^\0 \d{A} = 1$ and $\nabla^2 C^\1 = f(C^\0)$, $\vec{n}\cdot \nabla C^\1|_{\Gamma_\t{cap}} = C^\1|_{\Gamma_\t{vil}} = 0$. In this case, the flux delivered to fetal blood yields the effective area {
\begin{equation}
\ell^2 = -\frac{1}{A_\t{cap}}\,\int_{\Gamma_\t{cap}}\! C^\1 \d{A}\,.  
\label{eq:weak-metabolism-FL}
\end{equation}}
The total uptake flux (in dimensional variables) is approximated by {
\begin{equation}
N_\t{FL} \approx B\,Q\,c_\t{mat}\left[1 - \frac{q_\t{max}}{D_\t{t}c_\t{mat}}\,\ell^2 \right]\,, 
\label{eq:N_FL}
\end{equation}}
or equivalently (using \eqref{eq:FL-flux-with-G(alpha)}) {$G\approx 1-\varepsilon \ell^2$.  The extreme flow-limited assumption requires that $N_{FL}$ is substantially smaller than the flux entering villous tissue, which balances the overall rate of uptake $\mathcal{V} q_{\max}$.}}


\chgrev{The limiting cases \eqref{eq:N_DL} and \eqref{eq:N_FL} provide an approximate ratio \chgrev{
\begin{equation}
\frac{F}{G} \approx 1 + \mathcal{W}\, \left[ 1 - \frac{\ell^2_{DL}}{\ell^2} \right]\,, \quad
\mathcal{W} \equiv \frac{q_\t{max}}{D_\t{t}\,c_\t{mat}}\,\ell^2\,. 
\label{eq:F-G-ratio-weak-metab}
\end{equation}}
As Table~\ref{tab:geometric_parameters} illustrates,  $\ell^2_{DL}/\ell^2$ for Specimens 1-4 
in all cases is less than $12$\%, indicating that uptake has a stronger relative effect on the flow-limited compared to the diffusion-limited state.  Thus, to a good approximation, we can describe the boundary between flow-limited and diffusion-limited uptake using $\mathrm{Da}^{-1}\approx 1+\mathcal{W}$ when uptake is weak (see inset to Fig.~\ref{fig3:region-diagram}), highlighting $\mathcal{W}=\alpha \ell^2/D_t$ as a significant dimensionless measure of uptake. }

\chgrev{The condition $N_{FL}\ll \mathcal{V}q_{\max}$ underpinning the flow-limited approximation can be expressed for first-order kinetics, using (\ref{eq:N_FL}), as $\mathrm{Da}^{-1} \ll (\mathcal{V}/\mathcal{L}\ell^2) \mathcal{W}$.  The geometric index $\mathcal{V}/\mathcal{L}\ell^2$ is an order unity parameter for all specimens (Table~\ref{tab:geometric_parameters}).  The condition $\mathrm{Da}^{-1}\sim \mathcal{W}\ll 1$ therefore provides an estimate of the conditions at which the flux partition ratio in Fig.~\ref{fig4:nonlinear-metabolism}B first falls appreciably below unity.}

\end{document}